\documentclass[aps,prd,twocolumn,showpacs,superscriptaddress,groupedaddress]{revtex4-1} 

\usepackage[utf8]{inputenc}
\usepackage[T1]{fontenc}
\usepackage{amsmath}    
\usepackage{amssymb}   
\usepackage{graphicx}   
\usepackage{verbatim}   
\usepackage{subfigure}  
\usepackage{hyperref}   
\usepackage{times}
\usepackage{footnote}
\usepackage{fixmath}
\usepackage{multirow}
\usepackage{epsfig} 
\usepackage{braket}
\usepackage{rotating}
\usepackage{float}
\begin{document}

\def\cb{\overline{c}}
\def\cbw{\cb^w}
\def\kb{\overline{k}}
\def\mn{{\mu\nu}}
\def\ct{\tilde{c}}
\newcommand{\av}[1]{\langle #1\rangle}

\def\al{\alpha}
\def\be{\beta}
\def\ga{\gamma}
\def\de{\delta}
\def\ep{\epsilon}
\def\ve{\varepsilon}
\def\ze{\zeta}
\def\et{\eta}
\def\th{\theta}
\def\vt{\vartheta}
\def\io{\iota}
\def\ka{\kappa}
\def\la{\lambda}
\def\vpi{\varpi}
\def\rh{\rho}
\def\vr{\varrho}
\def\si{\sigma}
\def\vs{\varsigma}
\def\ta{\tau}
\def\up{\upsilon}
\def\ph{\phi}
\def\vp{\varphi}
\def\ch{\chi}
\def\ps{\psi}
\def\om{\omega}
\def\Ga{\Gamma}
\def\De{\Delta}
\def\Th{\Theta}
\def\La{\Lambda}
\def\Si{\Sigma}
\def\Up{\Upsilon}
\def\Ph{\Phi}
\def\Ps{\Psi}
\def\Om{\Omega}
\def\cA{{\cal A}}
\def\cB{{\cal B}}
\def\cC{{\cal C}}
\def\cE{{\cal E}}
\def\cl{{\cal L}}
\def\cL{{\cal L}}
\def\cO{{\cal O}}
\def\cP{{\cal P}}
\def\cR{{\cal R}}
\def\cV{{\cal V}}
\def\mn{{\mu\nu}}

\def\fr#1#2{{{#1} \over {#2}}}
\def\half{{\textstyle{1\over 2}}}
\def\quar{{\textstyle{1\over 4}}}
\def\eigh{{\textstyle{1\over 8}}}
\def\frac#1#2{{\textstyle{{#1}\over {#2}}}}

\def\vev#1{\langle {#1}\rangle}
\def\expect#1{\langle{#1}\rangle}
\def\norm#1{\left\|{#1}\right\|}
\def\abs#1{\left|{#1}\right|}

\def\lsim{\mathrel{\rlap{\lower4pt\hbox{\hskip1pt$\sim$}}
    \raise1pt\hbox{$<$}}}
\def\gsim{\mathrel{\rlap{\lower4pt\hbox{\hskip1pt$\sim$}}
    \raise1pt\hbox{$>$}}}
\def\sqr#1#2{{\vcenter{\vbox{\hrule height.#2pt
         \hbox{\vrule width.#2pt height#1pt \kern#1pt
         \vrule width.#2pt}
         \hrule height.#2pt}}}}
\def\square{\mathchoice\sqr66\sqr66\sqr{2.1}3\sqr{1.5}3}

\def\prt{\partial}

\def\etal{{\it et al.}}

\def\pt#1{\phantom{#1}}
\def\ni{\noindent}
\def\ol#1{\overline{#1}}

\def\nsc#1#2#3{\om_{#1}^{{\pt{#1}}#2#3}}
\def\lsc#1#2#3{\om_{#1#2#3}}
\def\usc#1#2#3{\om^{#1#2#3}}
\def\lulsc#1#2#3{\om_{#1\pt{#2}#3}^{{\pt{#1}}#2}}

\def\tor#1#2#3{T^{#1}_{{\pt{#1}}#2#3}}

\def\vb#1#2{e_{#1}^{{\pt{#1}}#2}}
\def\ivb#1#2{e^{#1}_{{\pt{#1}}#2}}
\def\uvb#1#2{e^{#1#2}}
\def\lvb#1#2{e_{#1#2}}
\def\etul#1#2{\et^{#1}_{{\pt{#1}}#2}}
\def\hul#1#2{h^{#1}_{{\pt{#1}}#2}}

\def\ab{{\color{red} \overline{a}{}}}
\def\bb{\overline{b}{}}
\def\cb{{\color{red} \overline{c}{}}}
\def\db{\overline{d}{}}
\def\eb{\overline{e}{}}
\def\fb{\overline{f}{}}
\def\gb{\overline{g}{}}
\def\Hb{\overline{H}{}}

\def\kb{\overline{k}{}}
\def\pb{\overline{p}{}}
\def\sb{{\color{red} \overline{s}{}}}
\def\kfb{(\overline{k_F})}

\def\dep{\de p}

\def\psb{\overline{\ps}{}}

\def\twiddle{\lower4pt\hbox{\hskip-0pt{$\widetilde{}$}}}
\def\m@th{\mathsurround=0pt}
\def\cmapstochar{\mathrel{\rlap{
  \lower0.1pt\hbox{\hskip-1.75pt{$\mapstochar$}}}
  \raise0pt\hbox{\hskip2.5pt{$\twiddle$}}}}
\def\notsimfill{$\m@th\cmapstochar$}
\def\scroodle#1{\vbox{\ialign{##\crcr\notsimfill\crcr
  \noalign{\kern-4pt\nointerlineskip}
   $\hfil\displaystyle{#1}\hfil$\crcr}}}

\def\cmapstocharbig{\mathrel{\rlap{
  \lower0.1pt\hbox{\hskip0.25pt{$\mapstochar$}}}
  \raise0pt\hbox{\hskip4.5pt{$\twiddle$}}}}
\def\notsimfillbig{$\m@th\cmapstocharbig$}
\def\scroodlebig#1{\vbox{\ialign{##\crcr\notsimfillbig\crcr
  \noalign{\kern-4pt\nointerlineskip}
   $\hfil\displaystyle{#1}\hfil$\crcr}}}

\def\Btw{\scroodlebig{B}}

\def\X{t_{\la \mn \ldots}}
\def\Xupper{t^{\la \mn \ldots}}
\def\Xb{\overline{t}_{\la \mn \ldots}}
\def\Xbupper{\overline{t}^{\la \mn \ldots}}
\def\Xtw{\scroodle{t}_{\la \mn \ldots}}
\def\Btw{\scroodlebig{B}}

\def\Xtildew{\widetilde{t}^{\la \mn \ldots}}
\def\Xtildewl{\widetilde{t}_{\la \mn \ldots}}
\def\Xtwo#1#2{\scroodle{t}{}^{(#1,#2)}_{\la \mn \ldots}}

\def\mt{m^{\rm T}}
\def\ms{m^{\rm S}}
\def\mb{m^{\rm{B}}}
\def\mbp{m^{\prime\rm{B}}}
\def\qt{q^{\rm T}}

\def\af{(a_{\rm{eff}})}
\def\afp{(a^{\prime{\rm B}}_{\rm{eff}})}
\def\afB{(a^{\rm B}_{\rm{eff}})}
\def\cuB{(c^{\rm B})}
\def\aft{(a^{\rm T}_{\rm{eff}})}
\def\afs{(a^{\rm S}_{\rm{eff}})}

\def\afb{(\ab_{\rm{eff}})}
\def\afbb{(\ab^{\rm B}_{\rm{eff}})}
\def\afbp{(\ab^{\prime{\rm B}}_{\rm{eff}})}
\def\afbnp{\ab_{\rm{eff}}}

\def\abt{(\ab^{\rm T}_{\rm{eff}})}
\def\abs{(\ab^{\rm S}_{\rm{eff}})}

\def\afbm{(\ab^\mu_{\rm{eff}})}
\def\cbm{(\cb^\mu)}

\def\afbx#1{(\ab^{#1}_{\rm{eff}})}
\def\cbx#1{(\cb^{#1})}

\def\abw{(\ab^w)}
\def\abe{(\ab^e)}
\def\abp{(\ab^p)}
\def\abn{(\ab^n)}
\def\afbe{\afbx{e}}
\def\afbp{\afbx{p}}
\def\afbn{\afbx{n}}

\def\cbe{(\cb^e)}
\def\cbp{(\cb^p)}
\def\cbn{(\cb^n)}

\def\cbw{\cbx{w}}
\def\afbw{\afbx{w}}
\def\afw{(a^w_{\rm{eff}})}
\def\cw{(c^w)}

\def\cbpr{(c^{\prime{\rm B}})}
\def\cot{(c^{\rm T})}
\def\cs{(c^{\rm S})}

\def\cbt{(\cb^{\rm T})}
\def\cbs{(\cb^{\rm S})}
\def\cbb{(\cb^{\rm B})}
\def\cbbp{(\cb^{\prime{\rm B}})}

\def\hlv{h^{(1,1)}}
\def\hlnv{h^{(0,1)}}

\def\atw{\scroodle{a}{}}
\def\btw{\scroodle{b}{}}
\def\ctw{\scroodle{c}{}}
\def\ctwb{(\ctw{}^{\rm B})}
\def\dtw{\scroodle{d}{}}
\def\etw{\scroodle{e}{}}
\def\ftw{\scroodle{f}{}}
\def\gtw{\scroodle{g}{}}
\def\Htw{\scroodle{H}{}}

\def\aftw{(\scroodle{a}_{\rm{eff}})}
\def\aftwb{(\scroodle{a}{}^{\rm B}_{\rm{eff}})}
\def\atwt{(\scroodle{a}{}^{\rm T}_{\rm{eff}})}

\def\ctwt{(\ctw{}^{\rm T})}

\def\aloo{\al_{(0,0)}}
\def\alol{\al_{(0,1)}}
\def\algo{\al_{(1,0)}}
\def\algl{\al_{(1,1)}}

\def\Phm{\frac{\Ph}{m_S}}
\def\s{\pt{a}}

\def\zon{\ze_1}
\def\ztw{\ze_2}
\def\zes{\ze_1^{\rm S}}
\def\zet{\ze_1^{\rm T}}
\def\zts{\ze_2^{\rm S}}
\def\ztt{\ze_2^{\rm T}}

\def\xc{\si_1}

\def\vf{j}

\def\noe{N_\oplus}

\def\axw{\afbw_{X}}
\def\ayw{\afbw_{Y}}
\def\azw{\afbw_{Z}}
\def\akw{\afbw_{K}}
\def\cxxw{(\cb^w)_{XX}}
\def\cyyw{(\cb^w)_{YY}}
\def\czzw{(\cb^w)_{ZZ}}
\def\cxyw{(\cb^w)_{(XY)}}
\def\cxzw{(\cb^w)_{(XZ)}}
\def\cyzw{(\cb^w)_{(YZ)}}
\def\ctxw{(\cb^w)_{(TX)}}
\def\ctyw{(\cb^w)_{(TY)}}
\def\ctzw{(\cb^w)_{(TZ)}}
\def\cttw{(\cb^w)_{TT}}
\def\ctjw{(\cb^w)_{(TJ)}}

\def\cqb{\cb_Q}
\def\cqbx#1{(\cb^{#1})_Q}

\def\cxx{\cb_{XX}}
\def\cyy{\cb_{YY}}
\def\czz{\cb_{ZZ}}
\def\cxy{\cb_{(XY)}}
\def\cxz{\cb_{(XZ)}}
\def\cyz{\cb_{(YZ)}}
\def\ctx{\cb_{(TX)}}
\def\cty{\cb_{(TY)}}
\def\ctz{\cb_{(TZ)}}

\def\stt{\sb_{TT}}
\def\sxx{\sb_{XX}}
\def\syy{\sb_{YY}}
\def\szz{\sb_{ZZ}}
\def\sxy{\sb_{(XY)}}
\def\sxz{\sb_{(XZ)}}
\def\syz{\sb_{(YZ)}}
\def\stx{\sb_{(TX)}}
\def\sty{\sb_{(TY)}}
\def\stz{\sb_{(TZ)}}
\def\that{{\hat{t}}}
\def\xhat{{\hat{x}}}
\def\yhat{{\hat{y}}}
\def\zhat{{\hat{z}}}
\def\jhat{{\hat{j}}}
\def\khat{{\hat{k}}}
\def\lhat{{\hat{l}}}
\def\muhat{{\hat{\mu}}}
\def\mnhat{{\hat{\mu}}{\hat{\nu}}}

\def\jearth{{\tilde{j}}}
\def\kearth{{\tilde{k}}}
\def\learth{{\tilde{l}}}
\def\tearth{{\tilde{t}}}
\def\xearth{{\tilde{x}}}
\def\yearth{{\tilde{y}}}
\def\zearth{{\tilde{z}}}
\def\muearth{{\tilde{\mu}}}
\def\mnearth{{\tilde{\mu}}{\tilde{\nu}}}

\def\acc{{\rm a}}

\def\lrpartial{\raise 1pt\hbox{$\stackrel\leftrightarrow\partial$}}
\def\lrprt{\stackrel{\leftrightarrow}{\partial}}
\def\lrprtnu{\stackrel{\leftrightarrow}{\partial^\nu}}
\def\lrDmu{\stackrel{\leftrightarrow}{D_\mu}}
\def\lrDnu{\stackrel{\leftrightarrow}{D^\nu}}
\def\lrvec#1{\stackrel{\leftrightarrow}{#1} }

\def\a{$a_\mu$}
\def\b{$b_\mu$}
\def\c{$c_{\mu\nu}$}
\def\d{$d_{\mu\nu}$}
\def\e{$e_\mu$}
\def\f{$f_\mu$}
\def\g{$g_{\la\mu\nu}$}
\def\H{$H_{\mu\nu}$}

\def\G{G_N}

\def\hnr{H_{\rm NR}}
\def\hnrs#1{H_{{\rm NR},#1}}

\def\pb{\overline{p}}
\def\nb{\overline{n}}

\def\csk{\xi}
\def\cskc{\xi_{\rm clock}}

\def\generala{I}
\def\laba{II}
\def\labb{III}
\def\labc{IV}
\def\labe{V}
\def\labf{VI}
\def\labg{VII}
\def\labh{VIII}
\def\spacea{IX}
\def\spaceb{X}
\def\spacec{XI}
\def\llra{XII}
\def\photona{XIII}
\def\summarya{XIV}
\def\summaryb{XV}

\newcommand{\beq}{\begin{equation}}
\newcommand{\eeq}{\end{equation}}
\newcommand{\bea}{\begin{eqnarray}}
\newcommand{\eea}{\end{eqnarray}}
\newcommand{\bit}{\begin{itemize}}
\newcommand{\eit}{\end{itemize}}
\newcommand{\rf}[1]{(\ref{#1})}

\def\pno#1{PNO(#1)}

\def\np{\nu_P}
\def\ne{\nu_E}

\def\epcombo{e+p}
\def\pen{e+p-n}

\def\nwr#1{n^w_{#1}} 

\def\cb{\overline{c}}
\def\cbw{\cb^w}
\def\kb{\overline{k}}
\def\mn{{\mu\nu}}
\def\ct{\tilde{c}}

\title{Lorentz-symmetry test at Planck-scale suppression with nucleons in a spin-polarized $^{133}$Cs cold atom clock} 

\author{ H. Pihan-Le Bars$^1$, C. Guerlin$^{2,1}$, R.-D. Lasseri$^3$, J.-P. Ebran$^4$, Q. G. Bailey$^5$, S. Bize$^1$, E. Khan$^3$, P. Wolf$^1$}
\affiliation{$^1$SYRTE, Observatoire de Paris, PSL Research University, CNRS, Sorbonne Universités, UPMC
Univ. Paris 06, LNE, 75014 Paris, France}
\affiliation{$^2$Laboratoire Kastler Brossel, ENS-PSL Research University, CNRS, UPMC-Sorbonne Universités,
Collège de France, 75005 Paris, France}
\affiliation{$^3$Institut de Physique Nucl\'eaire, Universit\'e Paris-Sud, IN2P3-CNRS, 
F-91406 Orsay Cedex, France}
\affiliation{$^4$CEA, DAM, DIF, F-91297 Arpajon, France}
\affiliation{$^5$Embry-Riddle Aeronautical University, 3700 Willow Creek Road, Prescott, Arizona 86301, USA}

\begin{abstract}
We introduce an improved  model that links the frequency shift of the $^{133}\text{Cs}$ hyperfine Zeeman transitions $\ket{F = 3, m_F} \longleftrightarrow \ket{F = 4, m_F  }$ to the Lorentz-violating Standard-Model Extension (SME) coefficients of the proton and neutron. The new model uses Lorentz transformations developed to second order in boost and additionally takes the nuclear structure into account, beyond the simple Schmidt model used previously in SME analyses, thereby providing access to both proton and neutron SME coefficients including the isotropic coefficient $\ct_{\emph{{\tiny  TT}}}$. Using this new model in a second analysis of the data delivered by the FO2 dual Cs/Rb fountain at Paris Observatory and previously analysed in \cite{Wolf2006}, we improve by up to 12 orders of magnitude the present maximum sensitivities \cite{Kosteleck`y2011} on the $\ct_{\emph{{\tiny  Q}}}$, $\ct_{\emph{{\tiny  TJ}}}$ and $\ct_{\emph{{\tiny  TT}}}$ coefficients for the neutron and on the $\ct_{\emph{{\tiny  Q}}}$ coefficient for the proton, reaching respectively $10^{-20}$, $10^{-17}$, $10^{-13}$ and $10^{-15}$ GeV.

\end{abstract}

\maketitle

\section{Introduction}
Our best current fundamental theories, 
General Relativity (GR) and the Standard Model of particle physics, 
are not expected to be valid at the Planck scale, 
where presumably a theory of quantum gravity holds. 
This, 
among other motivations, 
has given rise to the study of unified theories such as string theory, 
or theories of quantum gravity such as loop quantum gravity. 
The Planck scale energy $E_P$ is on the order of $10^{19}$~GeV, 
and the highest energy experiments or observations are well below this scale 
(ultra-high energy cosmic rays have energy lower than $10^{11}$~GeV). 
So testing these theories has been displaced to low-energy scales, 
where suppressed relics from Planck-scale physics may be observable, 
resulting in deviations from known physics.

There has been widespread interest in the last two decades in searching for such deviations, 
particularly among the so-called ``quantum gravity phenomenology'' \cite{Mattingly2005,Amelino-Camelia2013,Liberati2013,Will2014,Tasson2014b}. 
Of the possible deviations from known physics, 
a central one is the breaking of the continuous spacetime symmetries: Lorentz symmetry, which is the invariance under three rotations and three boost (shown to be also associated with the discrete Charge, Parity, Time-reversal (CPT) symmetry in \cite{Luders1954,Pauli1957,Greenberg2002}).  
Observable spontaneous breaking of these two symmetries could for example arise in string field theory, as was suggested nearly three decades ago \cite{Kostelecky1989,Kostelecky1991}.

A widely used effective field theory describing Lorentz Invariance Violations (LIV) and CPT violations 
is the Standard-Model Extension (SME) \cite{Colladay1997,Colladay1998,Kostelecky2004,Kosteleck`y2011}.
Under the conservative physical assumptions of energy-momentum conservation and observer Lorentz invariance, 
the SME introduces all possible Lorentz- and CPT-violating tensor operators in the Lagrange densities of the Standard Model (and General Relativity), 
parametrized by coefficients. These coefficients can be seen as background tensor fields that are constant in space-time on the scale of solar system experiments, and lead to a fundamentally different LIV approach than e.g. space or time varying scalar field approaches \cite{Damour1994,Kostelecky2003,Damour2010,Hees2016}.
They are allowed to be species-dependent, and vanish in the case of perfect Lorentz and CPT symmetry. As a test framework, the SME does not predict values of the coefficients.
However they are generally expected to be suppressed by a power of
$E/E_P$ increasing with the dimension of the associated LIV operator,
where $E$ is a cut-off energy. In \cite{Kostelecky1995} $E$ has been
taken as the electroweak energy ($E_{ew}\sim10^{2}$~GeV), leading to suppressions by a power of $10^{-17}$.

We focus here on the matter sector (electron, proton, neutron) of the minimal SME (mSME) which includes Lorentz violating operators of mass dimension 3 and 4 in the Lagrange density. Many of the coefficients are already constrained at or below their expected suppression \cite{Kosteleck`y2011}. Among still poorly constrained coefficients in this sector are however several components of the CPT-even, traceless and symmetric $\cb_{\mn}$ tensor, namely $\ct_{\emph{{\tiny  TT}}}$ for the proton and neutron ($10^{-11}$ GeV), $\ct_\emph{{\tiny TJ}}$ for the neutron ($10^{-5}$ GeV), and $\ct_\emph{{\tiny Q}}$ for the neutron ($10^{-14}$ GeV) \cite{Lo2016}, where indices $(T,X,Y,Z)$ refer to the coordinates in the Sun Centered Celestial Equatorial Frame. In this study, re-analyzing with a more complete model the data taken in \cite{Wolf2006} on spin-polarized transitions in a $^{133}$Cs fountain clock, we bring the bounds for all these coefficients below or near one Planck scale suppression (i.e. $10^{-17}$~GeV), thereby improving them by up to 12 orders of magnitude. We constrain for the first time independently all $\cb_{ \mn}$ components simultaneousy. We find no evidence for Lorentz violation, which challenges suppressions generally expected from quantum gravity phenomenology or helps setting limits on the cut-off energy $E$ (\cite{Collins2004,Mattingly2008, Pospelov2012,Alexandre2016}).

\section{Methods summary}\label{sec:method-summary}
Our generic approach is the following. LIV is manifested in our experiment as an anisotropy of the nucleons dispersion relation. Two states with a different nuclear momentum quadrupole moment undergo a different LIV energy shift, giving rise to a boost and orientation dependent shift of the transition frequency. We measure directly this frequency shift by interferometry on atomic wavefunctions, using the usual clock Ramsey interrogation sequence but applied on spin-polarized states.

A Lorentz transformation allows to express this lab LIV frequency shift in terms of the $\cb_{ \mn}$ coefficients in an inertial frame (usually taken for Solar System experiments as the Sun Centered Celestial Equatorial Frame, i.e. the solar system rest frame). It combines a rotation, which gives sensitivity to $\cb_{\emph{{\tiny JK}}}$ coefficients, and boosts, giving sensitivity to $\cb_\emph{{\tiny \tiny TJ}}$ (suppressed by one order in boost) and to the isotropic $\cb_\emph{{\tiny TT}}$ coefficient (suppressed by 2 orders in boost). The latter is unobservable in an inertial frame since it is only an overall rescaling of energies, but in a non uniformly boosted frame it gives rise to time variations at sidereal and annual frequency sidebands. LIV observables being usually expressed only to order one is boost, this coefficient has mostly been dismissed so far in this type of test, and is consequently among the less constrained coefficients of the $\cb_{ \mn}$ tensor. Its current best limits are set by its gravitational effects; for nucleons, torsion balance experiments bring a constraint on a linear combination involving this coefficient, at the $10^{-11}$~GeV level \cite{Kosteleck`y2011}. This gravitational LIV shift is negligible for hyperfine transitions, so without loss of generality the clock observable is derived here in flat space-time.

We use a set-up usually operated as a Cs fountain clock contributing to TAI in non-magnetic ($m_F = 0$) states. For testing SME, it has been operated on magnetized ($m_F \neq 0$) states to allow LIV tests during two periods respectively of 21 and 14 days at half a year interval. The two data sets have already been analyzed in \citep{Wolf2006} and led to 5 new independent constraints on 8 components of the proton $\cb_{ \mn}$ tensor. In this second analysis, our advanced mSME model allows us to disentangle the 9 components and provides new limits the isotropic component $\cb_\emph{{\tiny \tiny TT}}$ which was not included in the previous analysis.

We also introduce an alternative calculation of the nuclear quadrupole moment, also investigated in \citep{Brown2016} and \citep{Flambaum2016}, to address the shortcomings of the usual Schmidt shell model considered so far in the derivation of SME clock observables, which only takes into account a single nucleon contribution \citep{Kostelecky1999}, and therefore does not provide a realistic description of the nucleus for most atoms. The calculations are performed with self-consistent relativistic mean field theory (SCRMF), which allows to go beyond the single nucleon model and to calculate both the neutron and proton contributions to the nuclear quadrupole moment involved in the SME LIV shift.

In Section \ref{sec:lab-frame} we recall the main features of the description of alkali hyperfine transitions in the SME and our experimental set-up. This section is kept brief as details are known in the literature and can be found in the cited references. We then describe in Section \ref{sec:SCF} the transformation to the sun centred frame including the second order boost  and the resulting model for our experiment, with some of the details relegated to an Appendix. After a description of our data analysis in Section \ref{data_analysis}, we first present in Section \ref{res_schmidt} our results using the Schmidt nuclear model, as usual in SME analyses, in order to facilitate comparison with previous results from this and other experiments. We then briefly describe the SCRMF nuclear calculations (Section \ref{sec:alt-nuc}) and provide in Section \ref{sec:alt_res} our constraints based on this nuclear model. Section \ref{sect:discussion} is devoted to a general discussion with conclusions and perspectives in Section \ref{sect:conclusion}.

\section{SME frequency shift in the lab frame for alkali hyperfine transitions and experimental setup}\label{sec:lab-frame}

Treating in all generality the SME shift on an atomic transition includes: curved space time, 8 mSME tensors, and summing over all electrons and nucleons.
As shown previously (\cite{Kostelecky1999,Bluhm2003,Kosteleck`y2011a}), treated as a perturbation to usual hyperfine energy levels in alkali, LIV gives rise to a dipolar shift (parametrized in the SME by the $\bar{b}_\mu$, $\bar{d}_\mn$ and $\bar{g}_{\lambda\mn}$ tensors) and a quadrupolar shift (parametrized by the $\cb_\mn$ tensor). The dipolar SME shift of a $\ket{F,m_F}$ state is proportional to $m_F$. In the specific linear combination of transition frequencies used as the observable, following the approach of \cite{Wolf2006}, the linear dependence in $m_F$ of our observable is nullified in order to reduce systematic shifts from magnetic fields. Our test is thus sensitive only to the quadrupolar shift, which we detail hereunder.

As shown previously (\cite{Kostelecky1999,Bluhm2003,Kosteleck`y2011a}), hyperfine states of alkali atoms are affected by a LIV quadrupolar energy shift given in curved space-time by the expectation value of the operator $\sum_w\delta \hat{H}_w$ with $w = p,n,e$ for proton, neutron and electron, where for each particle:
\begin{equation}
\label{eq-deltaH}
\delta \hat{H}=\dfrac{2U}{3c^2}\cb_{\tiny tt}\dfrac{\hat{\mathbold{p}}^2}{2m}-\dfrac{1}{6m}{\cal C}^{(2)}_0 \hat{\cal
  P}^{(2)}_0.
\end{equation}
Here we omitted the $w$ index for the sake of simplicity. The
cartesian coordinate components are labeled with indices $(t,x,y,z)$
and relate to the space-time lab frame; the direction $z$ is taken
along the quantization axis. $U$ is the Newtonian gravitational
potential, $\hat{\mathbold{p}}$ and $m$ are respectively the momentum
operator and the mass of the particle. The spherical tensor component ${\cal T}^{(r)}_q$ associated with a tensor ${\cal T}_\mn$ is used here for a compact formulation, with $r$ its rank and $q\in(-r,..r)$ the index of its spherical components. The ${\cal T}^{(2)}_0$ component appearing here is linked to the cartesian coordinates via ${\cal T}^{(2)}_0={\cal T}_{jj}-3{\cal T}_{zz}$ with the convention of summation over like indices. This notation is used both for tensors ${\cal C}_{ij}=\cb_{ij}$ and $\hat{{\cal P}}_{ij}=\hat{p}_i\hat{p}_j$: 
\begin{eqnarray}
{\cal C}^{(2)}_{0}=\cb_{jj}-3\cb_{\tiny{zz}}\\
\hat{{\cal P}}^{(2)}_0=\hat{{\mathbold p}}^2-3\hat{p}_z^2.
\end{eqnarray}

The first term of the quadrupolar LIV operator in Eq. \ref{eq-deltaH} leads to an anomalous gravitational redshift \cite{Kosteleck`y2011a}. It has been used in the analysis of the spectroscopy of an electronic transition in Dysprosium to provide a gravitational constraint on the electron $\cb_\emph{{\tiny TT}}$ coefficient \cite{Hohensee2013}. This anomalous redshift scales as the differential internal kinetic energy between two states; therefore it is relevant for electronic transitions, but it is negligible for transitions between hyperfine states, which differ essentially via the relative orientation of the nuclear and electronic spin. So this contribution plays a negligible role and will be ignored in the following.

The second term scales with
the quadrupole moment operator of the momentum, and can be regarded as an anisotropy of each particle's kinetic energy. It is governed by the quadrupole moment of the $\cb_\mn$ tensor, which is usually, in the minimal SME, expressed in terms of energy for each particle by 
\begin{equation} \ct_q=mc^2{\cal C}^{(2)}_{0}.\end{equation}

In an atomic hyperfine state $\ket{F,m_F}$, the perturbative energy shift contribution from each particle is therefore proportional to the expectation value of the momentum quadrupole moment operator $\bra{F,m_F}\hat{{\cal P}}^{(2)}_0\ket{F,m_F}$, which using the Wigner-Eckart theorem can be expressed as a function of the expectation value in the extremal $m_F=F$ state $\bra{F,F}\hat{{\cal P}}^{(2)}_0\ket{F,F}$ with a prefactor $\hat{m}_F$ (following the notation of \cite{Kostelecky1999}):
\begin{eqnarray}
\hat{m}_F&=& \dfrac{\bra{F,m_F}\hat{{\cal F}}^{(2)}_0\ket{F,m_F}}{\bra{F,F}\hat{{\cal F}}^{(2)}_0\ket{F,F}}\\
& =&\dfrac{F(F+1)-3m_F^2}{F(F+1)-3F^2}.
\end{eqnarray}
This is a Clebsch-Gordan coefficient involving the quadrupole moment operator $\hat{{\cal F}}^{(2)}_0=\hat{{\mathbold F}}^2-3\hat{F}_z^2$ of the tensor $\hat{F}_i\hat{F}_j$ with $\mathbold{F}$ the total magnetic moment of the atomic state; it does not depend on the considered particle within the atom.

Summing from Eq. \ref{eq-deltaH} over all particles, the total LIV perturbative energy shift of a state $\ket{F,m_F}$ can therefore be expressed as:
\begin{equation}
\label{eq-DeltaE}
\delta E=\hat{m}_F\sum\limits_w\gamma^w\ct_q^w
\end{equation}
with particle dependent dimensionless scaling factors
\begin{equation}
\gamma^w=\dfrac{M_{20}^w}{6m_w^2c^2}
\label{eq-G}
\end{equation}
wich are proportional to the total momentum quadrupole moment from the $N_w$ particles of type $w$:
\begin{equation}
M_{20}^w=-\sum\limits_{N=1}^{N_w}\bra{F,F}\hat{\cal P}^{(2)}_{0,w,N}\ket{F,F}.
\label{eq-M}
\end{equation}



\begin{figure}
\includegraphics[width = 0.9\linewidth]{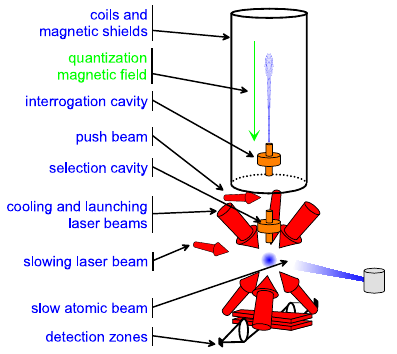}
\caption{Schematic view of an atomic fountain, from \citep{Bize2005}.}
\label{fig1}
\end{figure}

The analysis is performed using the data obtained from the $^{133}$Cs
and $^{87}$Rb dual fountain FO2 (see Figure \ref{fig1}),
operating at the Paris Observatory, already used in
\citep{Wolf2006}. An atomic gas is laser cooled and launched upwards
on a ballistic trajectory. The desired $\ket{F,m_F}$ initial state is
prepared in the selection cavity and remaining atoms in other states
are pushed away by a push beam. A microwave cavity allows us to perform a Ramsey interferometry sequence as the atoms pass through its mode successively during their upward and downward passage. For further experimental details, we refer the reader to an abundant litterature (e.g. \citep{Guena2012,Guena2014}).

Our data consist of two measurement sets of 21 and 14 days duration, taken respectively in April and September 2005. During the experiment, the fountain was run in Cs mode, which interrogates transitions between the $F=3$ and $F=4$ hyperfine levels of the ground state $6^2S_{1/2}$, with a magnetic field oscillating at 9.2~GHz. It was interrogated alternatively on three $\ket{ 3 , m_F } \longleftrightarrow \ket{ 4 , m_F }$ transitions with $m_F=+3$, $-3$ and $0$, of respective frequencies $\nu_{+3}$, $\nu_{-3}$ and $\nu_0$.

The quantization magnetic field is vertical. The rotational and orbital motion of the Earth then provides a change of the orientation of this axis, as well as of the laboratory boost, with respect to the SCF frame (see Section \ref{sec:SCF}). In the presence of LIV, this would result in a time variation of the particles kinetic energy anisotropy, and therefore in a time-varying quadrupole shift of the atomic level energies described by Eq. \ref{eq-DeltaE}.

From Eq. \ref{eq-DeltaE}, the frequency shift of a $\ket{  F = 3, m_F } \longleftrightarrow \ket{  F = 4, m_F  }$ transition can be calculated as the differential energy shift between the two levels \cite{Bluhm2003}. We use the following combination of hyperfine transition frequencies: $\nu_c=\nu_{+3}+\nu_{-3}-2\nu_0$ already used in \cite{Wolf2006}, which preserves the sensitivity to the quadrupolar LIV shift while canceling magnetic perturbations from the first order Zeeman effect. From Eq. \ref{eq-DeltaE}, it can be shown that the LIV frequency shift of this observable is (\cite{Bluhm2003}):
\begin{equation}
\label{eq-nuc} 
\delta\nu_c=-\dfrac{9}{7h}\left[\gamma^p\ct_q^p+\gamma^n\ct_q^n\right],
\end{equation}
with an overall scaling $9/7h=3.1\times10^{23}$~Hz/GeV for SME coefficients expressed in GeV. Here $\gamma^w$ is defined as in Eq. \ref{eq-G} but with the momentum quadrupole moment
expectation value $M^w_{20}$ in Eq. \ref{eq-M} taken in the $\ket{I, I}$ state.

It is not sensitive to the electron SME coefficients. Indeed closed shells do not contribute \cite{Kostelecky1999}, and the symmetry of the orbital of the valence electron is spherical since its orbital angular momentum in the state $6^2S_{1/2}$ is $L=0$; this leads a zero value of the momentum quadrupole moment in Eq. \ref{eq-M}. Our experiment thus offers a test of LIV sensitive only to nucleons, allowing to give constraints decorrelated from the electron LIV coefficients. The frequency shift is related to each nucleon SME coefficient through a dimensionless factor $\gamma^w$ via the nucleon momentum quadrupole moment $M_{20}^w$ (Eq. \ref{eq-G}) obtained from the nuclear model (Eq. \ref{eq-M}). The explicit dependence for each nucleon thus requires the choice of a nuclear model, for which there is no simple description for heavy nuclei such as $^{133}\mathrm{Cs}$. We adress this question further, in Sec. \ref{res_schmidt} and \ref{sec:alt-nuc}. In the rest of the paper,
all references to notations $\gamma^w$, $M^w_{20}$, and to Eqs. \ref{eq-G} and \ref{eq-M} are
meant, as in Eq. \ref{eq-nuc}, with the momentum quadrupole
moment expectation value $M^w_{20}$ taken in the $\ket{I, I}$ state.

\section{Time variation in the Sun Centered Frame with second order boost} \label{sec:SCF}

As the lab frame rotates and moves around Earth and Sun, periodic modulations of the clock frequency appear when each lab frame SME coefficient $\ct_q^w$ in Eq. \ref{eq-nuc} is expressed in terms of the Sun Centered Frame (SCF) SME coefficients. The latter are supposed constant at the time scale of our experiment, since the SCF has a rectilinear uniform motion with respect to a cosmological frame like the one given by the Cosmic Microwave Background. The SCF celestial frame, defined in \citep{Kostelecky2002}, is conventionally used to report SME results and to compare them. If  ${\mathfrak T}$ is the Lorentz transformation from the SCF to the lab frame for a co-vector, the tensor component $\cb_{ij}$ in the lab frame is the following combination of the tensor components in the SCF:
\begin{equation}\label{equ:transfo}
\cb_{ij} = {{\mathfrak T}_i}^{\Pi}{ {\mathfrak T}_j}^{\Gamma} \cb_{\Pi\Gamma}
\end{equation}
where $\Pi, \Gamma$ are indices spanning the SCF coordinates $(T,X,Y,Z)$. As described in \citep{Bluhm2003},  ${\mathfrak T}$ is the composition of a Lorentz boost $\Lambda$ followed by a rotation $R$. The Lorentz boost $\Lambda$ involves the Earth's orbital boost which varies with annual frequency $\mathrm{\Omega}$, and the lab boost due to the Earth's rotation varying at sidereal frequency $\omega$, with respective magnitude $\beta_\oplus\sim 10^{-4}$ and $\beta_l\sim 10^{-6}$ (ratio of velocity to speed of light). The rotation is due to Earth's rotation, at sidereal frequency. The generic expressions of $\Lambda$ and $ {\mathfrak T}$ are given respectively in Eq. \ref{equ:Lambda1} and Eq. \ref{equ:T} in Appendix \ref{model}.

In the previous analysis \citep{Wolf2006} (as well as in most previous litterature e.g., \cite{Kostelecky1999,Bluhm2003}), the boost matrix $\Lambda$ was approximated at first-order of its Taylor expansion in $\beta$, and included only the dependence in the Earth's orbital boost $\beta_\oplus$. This $O\left( \beta \right)$ model led to stringent constraints on 8 of the 9 independent components of $\cb_\mn$ of the proton, improving state-of-the-art constraints by up to 12 orders of magnitude. The ninth one, the isotropic coefficient $\cb_{\text{{\tiny  \emph{TT}}}}$, is suppressed by a factor  $\beta^2$ and thus did not appear in this first order model.

The main motivation of the present analysis is that the high improvement demonstrated by these first results, based on the intrinsic high sensitivity of cold atom clocks, can benefit also to the more suppressed $\cb_{\text{{\tiny  \emph{TT}}}}$ terms. Since the suppression arises from the pure timelike nature of the $\cb_{\text{{\tiny  \emph{TT}}}}$ coefficient, the constraint is expected to be less stringent but this coefficient is the less constrained from other non-gravitational experiments as well. 
In this analysis we developped an improved SME model of our observable, which, using a second order boost expression of Eq. \ref{eq-nuc} in the SCF, includes the $\cb_{\text{{\tiny  \emph{TT}}}}$ coefficient. This approach has also been used previously in spectroscopy in \cite{Hohensee2013} to constrain electron $\cb_\mn$ coefficients. Initially, our model contained all the terms up to $O\left( \beta^2 \right)$ for the 9 independent components of $\cb_\mn$, but as they do not provide any valuable contribution to the analysis they are neglected, except for the $\cb_{\text{{\tiny  \emph{TT}}}}$ coefficient.
More details on this derivation and a summarized version of the model can be found in Appendix \ref{model} (Table \ref{final}).

The Lorentz violating signal with this $O\left( \beta^2 \right)$ model includes sinusoidal variations at base frequencies $\{0,\omega,2\omega\}$ associated with sidebands at annual frequency $\mathrm{\Omega}$  as for the first order model and new sidebands at $\mathrm{2\Omega}$. It exhibits in total 13 frequency components (25 quadratures), instead of 3 frequency components (5 quadratures) for the previous analysis which did not include the annual frequency nor the second order terms.

SCF $\cb_\mn$ components appear as 9 observable combinations, which as usually in the SME are given rescaled by the rest mass energy of the particle and will be referred to in the following as \citep{Kosteleck`y2011} :
\begin{eqnarray} \label{equ:cQdef}
\tilde{c}_{\tiny\emph{Q}} &=&mc^2  (\cb_{{\tiny\emph{XX}}}+\cb_{\tiny\emph{YY}}-2\cb_{\tiny\emph{ZZ}})\nonumber \\
\tilde{c}_- &=&mc^2  (\cb_{{\tiny\emph{XX}}}-\cb_{\tiny\emph{YY}})\\
\tilde{c}_{\tiny\emph{J}} &=&mc^2  (\cb_{\tiny\emph{KL}} + \cb_{\tiny\emph{LK}}) \nonumber \\
\tilde{c}_{\tiny\emph{TJ}} &=&mc^2  ( \cb_{\tiny\emph{TJ}} + \cb_{\tiny\emph{JT}}) \nonumber\\
\tilde{c}_{\tiny\emph{TT}} &=&mc^2 \cb_{\tiny\emph{TT}} \nonumber
\end{eqnarray}
where $J,K,L$ are indices spanning spatial SCF coordinates $(X,Y,Z)$. The index $w$ referring to the flavor of the particle (proton or neutron) has been omitted here for the mass $m$ and the tensor components.

\section{Data analysis}\label{data_analysis} 
The data processing is performed using a weighted least-squares adjustment \footnote{Because the data are dominated by white noise (see \cite{Wolf2006}), ordinary least-square fitting is well adapted for our analysis} of the data used in \citep{Wolf2006} to the $O\left( \beta^2 \right)$ model. In our observable in Eq. \ref{eq-nuc}, $\cb_q^w$ is expressed in terms of the SCF SME coefficients as described in Section \ref{sec:SCF}, and  the flavour-dependent scaling factor $\gamma^w$ value is set from the considered nucleus model, as will be described in Sections \ref{res_schmidt} and \ref{sec:alt-nuc}. From our measurements of $\nu_c$ we then adjust directly all nine independent SCF combinations given in Eq. \ref{equ:cQdef}, without intermediate steps as e.g. on the Fourier basis. As the data show white noise behaviour (see \cite{Wolf2006}),  the least-squares method provides robust limits on SME coefficients. Our 3~mHz data standard deviation averages over the two data sets down to a resolution of 60~$\mu$Hz  on the amplitude of a sinusoidal deviation, which sets the bottom limit of statistical sensitivity of our test as in \cite{Wolf2006}.

The main systematic effects which are not already corrected in the clock data, as detailed in \citep{Guena2012}, are the Zeeman frequency shifts. 
For the frequency combination in Eq. \ref{eq-nuc}, the second order Zeeman effect, proportional to $B^2$ with $B$ the magnetic field, adds up to an overall shift of $-2$~mHz for our data. The variations of this term due to magnetic field fluctuations lie well below our frequency resolution; as in \cite{Wolf2006}, this shift is therefore considered as constant and the measured value of $\nu_c$ is corrected for this constant offset prior to the SME model adjustment. 

The first order Zeeman shift, proportional to $B$, is theoretically rejected in the combined observable $\nu_c$. However atoms with $m_F=+3$ and $m_F=-3$ follow slightly different trajectories
, which in the presence of magnetic field inhomogeneities results in incomplete cancellation of the first order Zeeman shift in $\nu_c$. The residual shift can be estimated using the time of flight (TOF) of the atoms in the fountain, measured periodically during the clock operation for each data set. As described in \citep{Wolf2006}, the TOF data exhibit a systematic difference of $158$ $\mu$s between the $m_F= +3$ and $m_F= -3$ atomic clouds; a Monte-Carlo simulation allows to constrain the corresponding residual first order Zeeman shift of $\nu_c$ to a conservative estimate of $0\pm25$~mHz \cite{Wolf2006}. 

As this dominant systematic effect is susceptible to vary with temperature at daily and annual frequency, it could mimick non-zero SME coefficients. To assess this systematic bias, we use  the upper limit of the above calibration to convert TOF data to worst-case frequency shift data, which we adjust with the $O\left( \beta^2 \right)$ model by weighted least-squares. We obtain an amplitude for each SME coefficient (noted $X_{\tiny\mathrm{TOF}}^i$) with a statistical uncertainty $\sigma_{\tiny\mathrm{TOF}}^i$. The obtained amplitude is an upper bound in absolute value; we therefore estimate the systematic bias at 0 and the systematic variance at $(X_{\tiny\mathrm{TOF}}^i)^2+(\sigma_{\tiny\mathrm{TOF}}^i)^2$ \footnote{For simplicity we neglect the cross term $2 X_{\tiny\mathrm{TOF}}^i \sigma_{{\tiny\mathrm{TOF}}}^i$ since the systematics fit results show only either highly significant values or highly nonsignificant values.}.

To obtain the total variance-covariance matrix of the estimated SME coefficients, we sum the statistical (clock data) and systematic (TOF data) matrix elements, in which the systematic variances are defined as above. The values and uncertainties for SME coefficients depend on the nucleus model considered, and will be presented for the Schmidt nuclear model in Section \ref{res_schmidt}, and using a more realistic nuclear model in Section \ref{sec:alt_res}. The detailed discussion is in Section \ref{sect:discussion}. 

\section{Results based on the Schmidt nuclear model}\label{res_schmidt}

From the sensitivity in frequency variation of our data set (Sec. \ref{data_analysis}), the corresponding sensitivity on SME coefficients depends on the value of the scaling factors $\gamma^{p,n}$ (Eq. \ref{eq-nuc}) which are proportional to the momentum quadrupole moment of the nucleons in the extremal state $\ket{F,F}$  as given by Eq. \ref{eq-G} and Eq. \ref{eq-M}. As an exact calculation is typically not feasible, different simplification levels can be used. A simplified description of the nucleus is given by the Schmidt model, which has so far been used in many cases when reporting SME constraints from atomic spectroscopy. It relies on a shell description of the nucleus. With an odd number of protons (55) and an even number of neutrons (78) in the nucleus of $^{133}$Cs atoms, a shell model leads to closed neutron shells and to a single valence proton. In the Schmidt description, this nucleon carries the entire magnetic moment of the nucleus which is involved in the hyperfine splitting. The expectation values in Eq. \ref{eq-M} which in principle have to be summed over all nucleons, reduce in this description to the expectation value for this single proton, and to an overall zero value from the neutrons \cite{Kostelecky1999}: $\gamma^n=0$. With this nucleus
model, the only constraints we can extract from our data are thus on proton coefficients. In the SME article on clocks \cite{Bluhm2003} and in the first data analysis \cite{Wolf2006}, the approximated value of $\gamma^p$ has been taken as $\gamma^p=-1.1\times10^{-3}$ leading to an expected maximum resolution on $\ct^p_\emph{{\tiny Q}}$ of approximately $2\times10^{-25}$~GeV using the frequency resolution given in Sec. \ref{data_analysis} and the conversion factor expressed in Sec. \ref{sec:lab-frame}.

\begin{figure}
\centering
\includegraphics[width= 0.9 \linewidth]{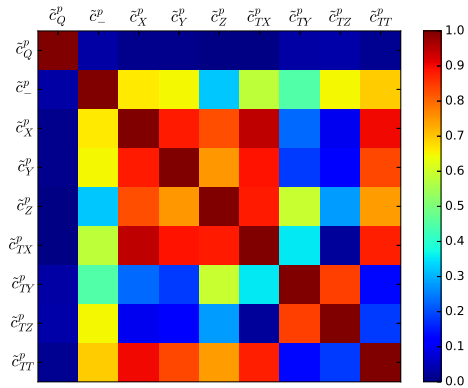}
\caption{Correlation matrix of the $\ct_\mn^p$ components. The matrix includes the statistic correlations (least-squares fitting of the data) and the systematic correlations (TOF), as described in Section \ref{data_analysis}.}
\label{fig2}
\end{figure}

Using these $\gamma^{p,n}$ values, our analysis provides the bounds on the $\ct_\mn^p$ components presented in Table \ref{coeff}. They are all consistent with Lorentz symmetry. The uncertainties show an improvement by 5 orders of magnitude on the $\ct^p_{\text{{\tiny  \emph{TT}}}}$ coefficient compared to the state-of-the-art constraints (\cite{Wolf2006,Kosteleck`y2011a}), reaching the $10^{-16}$ GeV scale. The correlation matrix is displayed on Figure \ref{fig2}. It contains high values, except for the $\ct^p_{\text{{\tiny  \emph{Q}}}}$ coefficient which is almost decorrelated at this sensitivity level. This indicates that the uncertainties contained in Table \ref{coeff} are marginalized uncertainties dominated by those correlations, and could thus be significantly improved with additional data spread over one year. 

\begin{table*}
\caption{Limits on SME Lorentz violating parameters $\ct_\mn^p$ for the proton, in GeV, when using the Schmidt model. The measured values and total uncertainties are shown together with the statistical (first bracket) and systematic (second bracket) uncertainties. These limits are obtained using a complete least-square adjustment of the $O\left( \beta^2 \right)$ model to the Cs fountain data. Constraints improved compared to state-of-the-art are displayed in bold, with in brackets the improvement factor in orders of magnitude. Note that previous constraints from \citep{Wolf2006} did not determine all coefficients independently (c.f. section \ref{pmvsc}).}
\centering
$\begin{array}{crrccc}
\hline \hline
\text{Coefficient} &  \multicolumn{2}{c}{ \text{Value and uncertainty}}    & \text{Unit (GeV)}& \text{Previous bound (GeV)} &\text{Ref.}\\
 \hline
\ct^p_{\mathrm{Q}}& -0.3 \pm 2.1 & (10^{-2})(2.1) &10^{-22} & 2.2  \enskip 10^{-22} & \text{\cite{Wolf2006}} \\
\ct^p_{-}&   1.4 \pm 9.0& ( 	0.7) ( 8.9) &10^{-24}	 & 2.8   \enskip 10^{-25}   &  \text{\cite{Wolf2006}} \\
\ct^p_\emph{{\tiny X}}&   -1.5 \pm 5.3&( 		0.7)( 	5.2)  &10^{-24}&  1.2  \enskip 10^{-25} & \text{ \cite{Wolf2006}}\\
\ct^p_\emph{{\tiny Y}}&  0.8 \pm  1.6	& ( 	  	0.3)( 		1.6 )  & 10^{-24}	& 	 1.2  \enskip 10^{-25} &  \text{\cite{Wolf2006} }\\
\ct^p_\emph{{\tiny Z}}&  	1.0  \pm 	3.9& ( 	  	0.8)( 		3.9 )	 &10^{-24}\ & 2.8  \enskip 10^{-25} &  \text{\cite{Wolf2006}} \\
\ct^p_\emph{{\tiny TX}}&  -1.5 \pm 5.7 &( 	 	0.6)( 	 	5.7)   & 10^{-20}	& 3.0  \enskip 10^{-21} & \text{ \cite{Wolf2006}}\\
\ct^p_\emph{{\tiny TY}}&   1.4	 \pm 5.9&( 		0.3)( 	 	5.9 )&10^{-20}  & 3.0  \enskip 10^{-21} &  \text{\cite{Wolf2006}} \\
\ct^p_\emph{{\tiny TZ}}&   -1.1 \pm	3.5& ( 	  	0.2)( 	 	3.5)  &10^{-20} & 2.0  \enskip  10^{-21} &  \text{\cite{Wolf2006}}\\
\ct^p_\emph{{\tiny TT}}& 1.6  \pm 	\mathbf{	6.9}& ( 	 	0.9)( 		6.9) & \mathbf{10^{-16 (5)}}   & < 10^{-11}& \text{ \cite{Kosteleck`y2011a}}\\
\hline
\hline
\end{array}$
\label{coeff}
\end{table*}

\begin{figure}
\begin{flushleft}
\hspace*{-5mm}\includegraphics[width =0.56\textwidth]{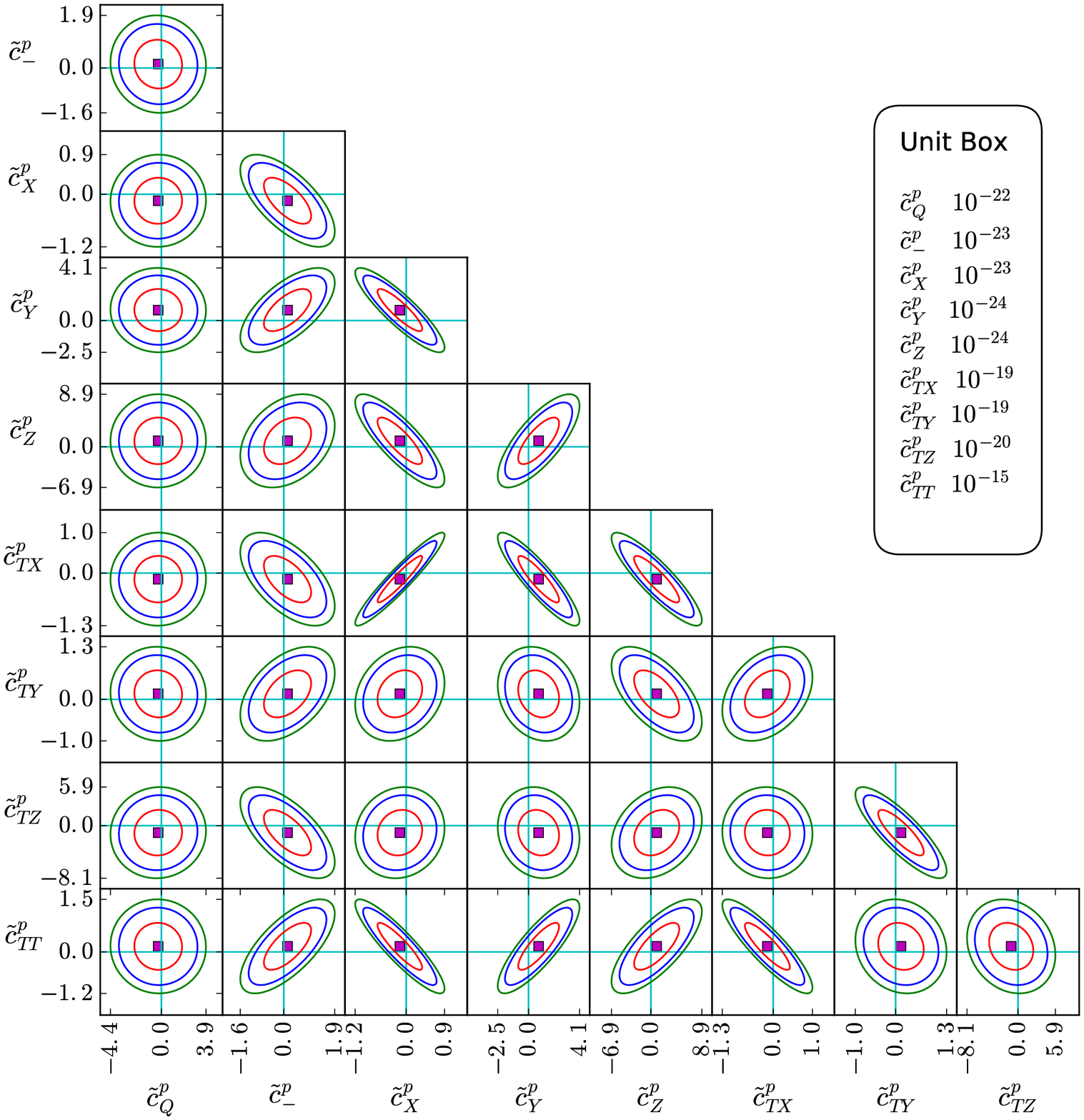}
\end{flushleft}
\caption{Confidence ellipses of the $\ct_\mn^p$ components, when using the Schmidt model. The red, blue and green confidence ellipses correspond respectively to the $ 68.3\%$, $90\%$ et $95.4\%$ confidence regions. The purple square is the position of the least-square solution. Axis labels give the $95.4\%$ confidence intervals in GeV, the respective orders of magnitude are given in the upper right box. Horizontal and vertical blue lines at 0 value allow to visualize the absence of significance in the results.}
\label{fig3}
\end{figure}

In addition to the correlation matrix, we provide confidence intervals on Figure \ref{fig3}, which allow a synthetic view over significance levels, orders of magnitude and correlations. We employ an analytical method developed in \citep{Brinker1995}, which uses the covariance matrix to build confidence ellipses whose semi-major axes are scaled by a given value of $\sqrt{\Delta \chi^2}$ \footnote{Here $\Delta \chi^2$ is the difference between the $\chi^2$ value at a given point in the parameter plane and the one of the best fit solution.} depending on the required probability. Some of the ellipses show a strongly diagonal orientation, indicating the presence of correlations between the SME coefficients that are in agreement with the correlations visible in Fig. \ref{fig2}.

In Appendix \ref{svd}, we provide an alternative, but entirely equivalent, description of the results under the form of independently constrained linear combinations of coefficients. These combinations are obtained from the Singular Value Decomposition (SVD) of the covariance matrix, meaning that they are the set of othonormal vectors (in the euclidean sense) that diagonalize the covariance matrix. Having no correlation between these combinations implies that the uncertainty is not degraded and thus reaches a lower value as can be seen in Table \ref{coeffsvd}, typically decreased by a factor of 2. The linear combinations given in Table \ref{composvd} have dominant coefficients; the corresponding constraints can thus be regarded as the ``maximal sensitivity'' on this coefficient from our data, in the sense commonly used when reporting SME constraints \cite{Kosteleck`y2011}.

\section{Self consistent relativistic mean field (SCRMF) nuclear model} \label{sec:alt-nuc}

\subsection{Relativistic Mean Field Formalism}
\label{rmf}
Recent developments in nuclear physics attracted a renewed interest in the effects of Lorentz-violation in atomic and nuclear physics \cite{Brown2016, Flambaum2016}. Following this trend we compute the nuclear matrix elements, in particular the ones required for the determination of $\gamma^w$ (Eq. \ref{eq-DeltaE}, \ref{eq-G} and \ref{eq-nuc}), in a fully microscopic way using a state-of-the-art nuclear structure approach.
This allows us to go beyond the single nucleon model and to calculate
both the neutron and proton contributions to the nuclear matrix
elements. The theoretical framework used here is the relativistic
energy density functional which is particularly suited to describe
nuclear structure properties in great depth \cite{dario-relat}.  In
this approach the nucleus is described in terms of nucleons considered
as point-like Dirac particles, while the interaction among them is
described by an exchange of mesons. Thus one may construct in a fully-covariant way a phenomenological
Lagrangian density which conserves the symmetries of the nuclear interaction:

\begin{multline}
\mathcal{L}= \bar{\psi} \big[ i \gamma^{\mu}\partial_{\mu} - m - g_{\sigma}\sigma - g_{\omega} \gamma_{\mu}\omega^{\mu} - g_{\rho} \gamma_{\mu} \vec{\rho}\cdot\vec{\tau}^{\mu} - \\ g_{\pi} \gamma_5\gamma_{\mu}\partial^{\mu}\vec{\pi}\cdot\vec{\tau} - e \gamma_{\mu} A^{\mu} \left( \dfrac{1-\tau_3}{2}\right)\big] \psi  + \mathcal{L}_k
\label{lagrangian}
\end{multline}
where $\mathcal{L}_k$ is the kinetic part of this Lagrangian. The
arrows symbolize vectors of the Isospin SU(2) space. $\psi$ is a Dirac
4-spinor describing a nucleon of mass $m$ while $\gamma_\mu$ denotes the usual Dirac matrices. In Eq. \ref{lagrangian} the
nucleons interact by the exchange of $\{ \sigma, \rho,\omega$ and
$\pi\}$ mesons. The coupling constants of \eqref{lagrangian} are then fitted in order to accurately reproduce the binding energies of a set of benchmarking nuclei \cite{tamara-solver}. 

\begin{figure}
\centering
\includegraphics[width= 1 \linewidth]{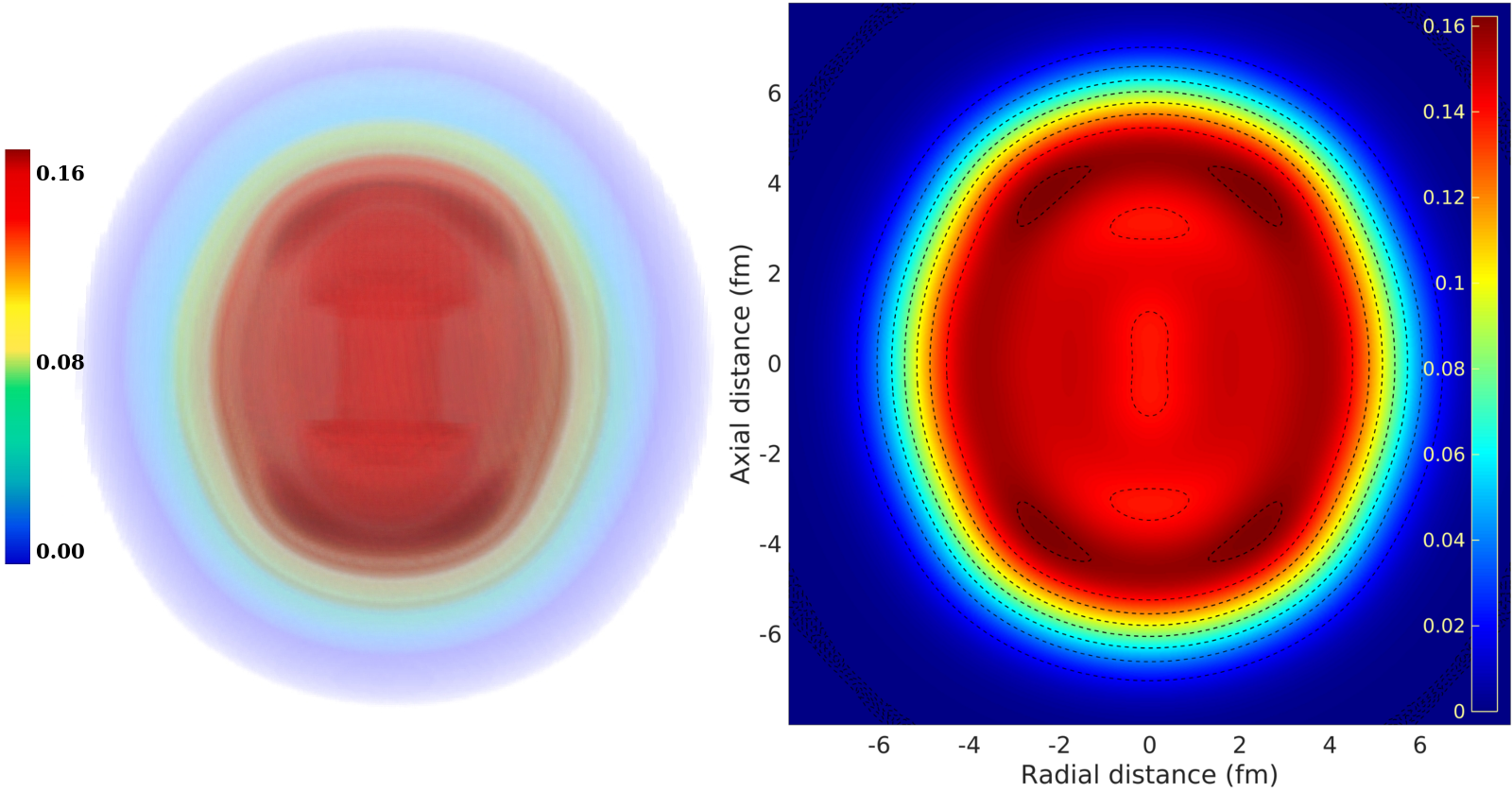}
\caption{Nucleonic density of $^{133}$Cs in fm${}^{-3}$, in the intrisic frame. Left: 3D representation. Right: 2D projection. The spatial scale is given in fm.}
\label{fig4}
\end{figure}

Performing a Legendre transform of \eqref{lagrangian} yields the Hamiltonian of the problem
\begin{multline}
\label{hamilton}
\mathcal{H} = \int d^3x \,\bar{\psi}[i \nabla + m ]\psi + \dfrac{1}{2}\int d^3x\, \bar{\psi} \big[ g_{\sigma} \sigma + g_{\omega} \gamma_{\mu}\omega^{\mu} \\ + g_{\rho} \gamma_{\mu} \vec{\rho}\cdot\vec{\tau}^{\mu}  + g_{\pi} \gamma_5\gamma_{\mu}\partial^{\mu}\vec{\pi}\cdot\vec{\tau} + e \gamma_{\mu} A^{\mu} \left( \dfrac{1-\tau_3}{2}\right)\big] \psi.
\end{multline}
The relativistic energy density functional is computed by taking the expectation value of \eqref{hamilton} on the vacuum state $\ket{\Phi_0}$
\begin{equation}
\label{efunc}
\mathcal{E}[\rho] = \braket{\Phi_0|\mathcal{H}|\Phi_0}
\end{equation}
while the density operator of the system is defined as:
\begin{equation}
\rho_{ij} = \dfrac{\braket{\Phi_0|c_j^{\dagger} c_i|\Phi_0}}{\braket{\Phi_0|\Phi_0}}.
\label{rho}
\end{equation}
with $c_i^{\dagger} / c_i$ being the nucleonic creation/anihiliation
operators. To compute the ground state energy, the variational
principle is applied to \eqref{efunc}. Neglecting the
Fock exchange term leads to the Relativistic Hartree-Bogoliubov
(RHB) equations, solved in a self-consistent way in an axially
deformed harmonic oscillator basis \cite{tamara-solver}.
For a given nucleus this model allows us to obtain its nuclear density in the nucleus intrinsic frame, as pictured on Fig. \ref{fig4} for $^{133}$Cs.


\subsection{Nuclear quadrupole moments}
The solutions of the RHB equations give access to the density of the
system. The moments of the multipolar expansion of the density
describe the shape of the nucleus. For instance a non-null quadrupole
moment in configuration space excludes spherical symmetry, but may
conserve an axial symmetry. The quadrupole moment in momentum space $M^w_{20}$ can be directly related to the $\gamma^w$ parameters by Eq. \ref{eq-G}.

In the
intrinsic frame of the nucleus the usual quadrupolar distribution is
expressed in configuration space as:

\begin{equation}
\tilde{Q}^w_{20} = \braket{\Phi_0|\left(3z^2 - r^2\right)|\Phi_0} =  \text{Tr}\left[\rho \left(3z^2 - r^2\right)\right].
\end{equation}
while in momentum space:
\begin{equation}
\label{p2value}
\tilde{M}^w_{20} =  \braket{\Phi_0|\left(3p_z^2 - \vec{p}\cdot\vec{p}\right)|\Phi_0}.
\end{equation}

In a final step we project the quadrupole moments into the laboratory frame using 
\begin{equation}
\label{projection}
Q^w_{20} = \dfrac{3K^2 - I(I+1)}{(I+1)(2I+3)} \tilde{Q}^w_{20}
\end{equation}
where $I$ is the nuclear spin of the considered nucleus and $K$ its projection onto the quantization axis. Note that \eqref{projection} gives only an approximate value in the laboratory frame. For better accuracy the complete projection of the RHB solutions on the correct total angular momentum needs to be performed, in the laboratory frame. This is out of the scope of the present paper but will be investigated in upcoming work using the Generator Coordinate Method (GCM) \cite{schuck}.

When comparing our results to the recently published ones in \cite{Flambaum2016}, which use a self-consistent mean field technique (SCMF), we note a good agreement (within a factor $\sim$1.3) for $M^p_{20}$ but very large discrepancies (up to a factor $\sim$60) for $Q^n_{20}$ and $M^n_{20}$. We ascribe this to the method used in \cite{Flambaum2016} which is tailored for the proton contribution as it is based on the experimental value of $Q^p_{20}$, and is thus likely to be only very approximate for the neutron quadrupole moments.

When carrying out our calculations for the atoms also studied in \cite{Brown2016} we find a very good agreement for $^{21}$Ne (for the protons $Q^p_{{\tiny SCRMF}} = 9.5 \text{ fm}^2$ when $Q^p_{{\tiny SCMF}}=9.7 \text{ fm}^2$ while the results are the same for neutron contribution)
and results of the same order of magnitude for $^{131}$Xe and $^{201}$Hg. The remaining differences are mainly due to the different methods used to constrain the computations.

In the case of $^{133}$Cs, the number of protons (Z = 55) is odd, thus
the spin-parity of the ground state is given by the valence
proton. In the framework of the energy density functional we use the usual half-filling approximation \cite{halffilling}, enforcing the correct spin-parity of the
state, here I=$\text{K}^{\small{\pi}}=\frac{7}{2}^{\small{+}}$. The comparison of the
experimental energy of the ground state ($E_{exp} = -1118.5$ MeV) to
the RHB predictions is given in Table \ref{tab:m}, showing the good
accuracy of the present approach. The quadrupolar moments in the
laboratory frame for $^{133}$Cs are given in Table \ref{tab:m} for
two different parametrizations of the relativistic functional \eqref{efunc} : DD-PC1
is a point-coupling interaction, while DD-ME2 takes into account the
full finite range meson exchange, and is more realistic.

\begin{table}[H]
\centering
\begin{tabular}{|p{2cm}|c|c|c|c|c|c|}
  \hline
  \quad Functional   & $Q^n_{20}$  &  $Q^p_{20}$ & $M^n_{20}$  &  $M^p_{20}$ & E (MeV)  \\
  \hline
  \quad DD-PC1 & -2.6576 & -0.3578 &  0.0047 & 0.1135 & -1118.6\\
  \quad DD-ME2 & -2.8083 & -0.3538 &  0.0024 & 0.1129 & -1117.7\\
  \hline   
\end{tabular}
\caption{Quadrupolar moments in configuration $\left(\text{fm}^2\right)$ and momentum representation $(\hbar/\text{fm})^2$ for $^{133}$Cs.}
\label{tab:m}
\end{table}

A more detailed presentation and discussion of these calculations for different nuclei as well as comparisons between different methods will be the subject of a future publication.

\section{Results based on the SCRMF nuclear model} \label{sec:alt_res}
From the two  nucleon interaction models considered in Sec. \ref{sec:alt-nuc}, we use the results of the DD-ME2 functional which are expected to be more realistic. From  Eq. \ref{eq-G} and Table \ref{tab:m} we obtain  $\gamma^p = 8.32\times 10^{-4}$ and $\gamma^n = 1.76\times 10^{-5}$. 
In this nuclear model the neutron contribution is not neglected unlike in the Schmidt model, so our experiment allows to constrain also neutron LIV, but with a sensitivity scaled by $\gamma^n$ about two orders of magnitude lower than for the proton.  We can not distinguish the relative nucleon contributions and are sensitive to the linear combinations $\ct_\mn^p+0.021\ct_\mn^n$. The limits on these SME coefficients combinations, given in Table \ref{coeff-SCRMF}, are obtained by a straightforward rescaling of Table \ref{coeff} from the Schmidt model, and consequently the correlation matrix stays the same as in Fig. \ref{fig2} with the proton coefficients replaced by the above combinations. Since the individual constraints from proton and neutron can not be disentangled, results for each nucleon can be expressed in terms of maximal sensitivity as defined in \cite{Kosteleck`y2011}. As shown in Table \ref{coeff-SCRMF}, improvements on proton coefficients are equivalent to the ones presented for the Schmidt model. For the neutron coefficients, our results improve by 12 orders of magnitude over state of the art for the $\ct_{\text{{\tiny \emph{TJ}}}}$ coefficients, 7 orders of magnitude for the $\ct_{\text{{\tiny \emph{Q}}}}$ coefficient, and 2 orders of magnitude for the $\ct_{\text{{\tiny \emph{TT}}}}$ coefficient, respectively down to $10^{-17}$, $10^{-20}$ and $10^{-13}$ GeV.


\begin{table*}
\caption{Limits (1 sigma) on SME Lorentz violating parameters $\ct_\mn^w$ for the proton and neutron, in GeV, when using SCRMF. The last two columns show the corresponding maximal sensitivities on each nucleon as defined in \cite{Kosteleck`y2011} (2 sigma limits logarithmically rounded). In bold are the values which improve over the state of the art published in the 2016 version of \cite{Kosteleck`y2011}, with in bracket the improvement in orders of magnitude \footnote{For the $\ct^n_\emph{{\tiny Q}}$ limit from state of the art we refer to the constraint given in the neutron sector part of \cite{Kosteleck`y2011} although it is not reported in the table summarizing maximal sensitivities in \cite{Kosteleck`y2011}.}.}
\centering
$\begin{array}{ccccc}
\hline \hline
\text{Coefficient} &  \text{Value and uncertainty}    & \text{Unit (GeV)} &  \ct^p_\text{max} \text{(GeV)}& \ct^n_\text{max}\text{(GeV)}\\
 \hline\vspace*{-3mm}\\
\ct^p_\emph{{\tiny Q}}+0.021\ \ct^n_\emph{{\tiny Q}}& 0.4 \pm 	2.8 & 10^{-22} &10^{-21} & \mathbf{10^{-20(7)}}  \vspace*{-.6mm}\\
\ct^p_- + 0.021\ \ct^n_- & -0.2 \pm 1.2 & 10^{-23} & 10^{-23} & 10^{-21}   \vspace*{-.6mm}\\
\ct^p_\emph{{\tiny X}} + 0.021\ \ct^n_\emph{{\tiny X}} &   2.0 \pm 7.0 &10^{-24}  &10^{-23}& 10^{-21} \vspace*{-.6mm}\\
\ct^p_\emph{{\tiny Y}} + 0.021\ \ct^n_\emph{{\tiny Y}}&  -1.1 \pm  2.2	& 10^{-24}  & 10^{-23}	& 	 10^{-22}  \vspace*{-.6mm}\\
\ct^p_\emph{{\tiny Z}} + 0.021\ \ct^n_\emph{{\tiny Z}}&  	-1.3  \pm 	5.2& 10^{-24}	 & 10^{-23} & 10^{-21}  \vspace*{-.6mm}\\
\ct^p_\emph{{\tiny TX}} + 0.021\ \ct^n_\emph{{\tiny TX}}&  2.0 \pm 7.6 &10^{-20}   & 10^{-19}	& \mathbf{10^{-17(12)}} \vspace*{-.6mm}\\
\ct^p_\emph{{\tiny TY}} + 0.021\ \ct^n_\emph{{\tiny TY}}&   -1.8	 \pm 7.8 & 10^{-20} & 10^{-19}  & \mathbf{10^{-17(12)}}  \vspace*{-.6mm}\\
\ct^p_\emph{{\tiny TZ}} + 0.021\ \ct^n_\emph{{\tiny TZ}}&  1.4 \pm	4.6& 10^{-21} &10^{-19} & \mathbf{10^{-17(12)}} \vspace*{-.6mm}\\
\ct^p_\emph{{\tiny TT}} + 0.021\ \ct^n_\emph{{\tiny TT}}& -2.2  \pm 9.1& 10^{-16} &\mathbf{10^{-15(4)}}   & \mathbf{10^{-13(2)}} \\
\hline
\hline
\end{array}$
\label{coeff-SCRMF}
\end{table*}

\section{Discussion} \label{sect:discussion}
\subsection{Comparison to previous works}
Our results show large improvements on the isotropic coefficient $\ct_{\text{{\tiny  \emph{TT}}}}$ for the proton with both nuclear models, as well as for several neutron coefficients for the relativistic nuclear model. With this model, by improving on the previous weakest limits our analysis brings all $\ct_\mn$ constraints for the proton and neutron below or much nearer to (for $\ct_{\text{{\tiny \emph{TT}}}}$) Planck scale suppression. All results presented in this paper are still consistent with Lorentz symmetry. 

For the proton, results are equivalent with both considered nuclear models. Our improved LIV model including annual modulations and terms at order $O\left(\beta^2\right)$ in boost leads to a high sensitivity to $\ct_{\text{{\tiny  \emph{TT}}}}$, that was not constrained by the previous analysis of our data.

The improvement on neutron coefficients comes from our new nuclear model that, unlike the Schmidt model, accounts for the sensitivity of our measurement to the neutron SME coefficients. The resulting limits on neutron coefficients are much less stringent than the ones from comagnetometers \citep{Smiciklas2011}. However, comagnetometer limits do not adress so far the boost dependent parts $\ct_{\text{{\tiny  \emph{TJ}}}}$ and $\ct_{\text{{\tiny  \emph{TT}}}}$ nor the spatial part $\ct_{\text{{\tiny  \emph{Q}}}}$, and this is where we provide large improvements. The previous $\ct_{\text{{\tiny  \emph{Q}}}}$ limit was set recently from acoustic waves in quartz \cite{Lo2016}.



In Sections \ref{res_schmidt} and \ref{sec:alt_res}, we compared our results on  $\ct_{\text{{\tiny  \emph{TT}}}}$ with those obtained through the SME WEP test interpretation of torsion balance experiments \citep{Kosteleck`y2011a}. This analysis however did not disentangle the isotropic component $\ct_{\text{{\tiny  \emph{TT}}}}$ from the spatial component $\ct_{\text{{\tiny  \emph{Q}}}}$; disentanglement has been done in \cite{Kosteleck`y2011a} and weakens this upper bound by 3 orders of magnitude. So in this respect the improvement factors of $10^4$ and $10^2$ displayed in Table \ref{coeff-SCRMF} are conservative, since when comparing to this disentangled limit we rather improve by 7 orders of magnitude the constraint on $\ct_{\text{{\tiny  \emph{TT}}}}$ for the proton, and by 5 orders of magnitude for the neutron.

\subsection{Role of the complete vs piecemeal analysis}
\label{pmvsc}

In our approach (which we refer to hereafter as ``complete'' analysis), the nine $\ct_\mn$ coefficients are fitted simultaneously. Fitting over a data set with sufficient resolution for sidereal period combined with a spread over half a year can enable us to discriminate all coefficients in a single fit, thanks to their contrasted spectral signatures. The uncertainty related to annual variation could however be strongly degraded by uncontrolled systematics. In our case, the robustness of the long term behaviour of clocks, which are built to provide absolute frequency references, gives us access to annual variation with controlled systematic uncertainty.

This simultaneous fit is in contrast to the most common approach, referred to hereafter as a  "piecemeal" analysis. In this approach, used e.g. in \citep{Wolf2006} and \citep{Hohensee2013}, the SME model adjustement is made successively on separate subsets of parameters assuming that they are independent and assigning in turn a zero expectation value for those not fitted. The drawbacks of this method are that it requires assumptions on the expectation values, and that by neglecting the correlation between parameters belonging to different subsets, it leads to an artificial decrease in the marginalized uncertainties and thereby to an underestimation of the individual parameter uncertainties. To illustrate this limit of the piecemeal analysis, we compared the results shown in \citep{Wolf2006}, obtained using such an analysis with two subsets of parameters, and the results obtained with a complete and direct fitting of the same first order $O\left( \beta\right)$ model to the data. For both the Schmidt model has been used.\par 

The results presented in Table \ref{min} show that, except for the $\ct^p_{\text{{\tiny  \emph{Q}}}}$ coefficient whose sensitivity is dominated by the systematics, the piecemeal analysis led to an underestimation of the uncertainties by a factor 6 to 20 in the previous analysis. That is why the constraints on the $\ct^p_\mn$ coefficients presented in Table \ref{coeff} show degraded uncertainties in comparison with \citep{Wolf2006}, except for $\ct^p_{\text{{\tiny  \emph{Q}}}}$, whose uncertainty remains the same and for $\ct^p_{\text{{\tiny  \emph{TT}}}}$ which was not constrained in \cite{Wolf2006}.

\begin{table}
\caption{Comparison between piecemeal (P) and complete (C) analysis for the $O\left( \beta \right)$ model used in \citep{Wolf2006}. In the complete analysis we fit directly for the SME coefficient values. For the piecemeal results we report the results obtained in \cite{Wolf2006} using as subsets the $\ct^p_{\text{{\tiny  \emph{TJ}}}}$  on the one hand and all the purely spatial combinations on the other hand. The underestimation factor of the uncertainty in the piecemeal analysis is denoted C/P.}
\centering
$\begin{array}{c c c c}
\hline \hline
\text{Coefficient}  &\text{Uncertainty (GeV) (P) \citep{Wolf2006}} & \text{Uncertainty (GeV) (C)} & \text{C/P} \\
 \hline\vspace*{-4mm}\\ 
\ct^p_{\text{{\tiny  \emph{Q}}}} & 2.2 \hspace*{1mm}10^{-22} &  2.1  \hspace*{1mm}10^{-22}& 0.95\\
\ct^p_{-}&   2.8 \hspace*{1mm}10^{-25} &  5.2 \hspace*{1mm}10^{-24} & 19\\
\ct^p_{\text{{\tiny  \emph{X}}}}&   1.2  \hspace*{1mm}10^{-25} &  1.2 \hspace*{1mm}10^{-24} & 10\\
\ct^p_{\text{{\tiny  \emph{Y}}}} &  1.2 \hspace*{1mm}10^{-25} & 7.5 \hspace*{1mm}10^{-25} & 6.3\\
\ct^p_{\text{{\tiny  \emph{Z}}}}&  2.8  \hspace*{1mm}10^{-25} & 2.8 \hspace*{1mm}10^{-24} & 10\\
\ct^p_{\text{{\tiny  \emph{TX}}}}&   3.0 \hspace*{1mm}10^{-21} &  2.3\hspace*{1mm}10^{-20}& 7.7\\
\ct^p_{\text{{\tiny  \emph{TY}}}}&  3.0 \hspace*{1mm}10^{-21} &  5.9 \hspace*{1mm}10^{-20}& 20\\
\ct^p_{\text{{\tiny  \emph{TZ}}}}&   2.0 \hspace*{1mm}10^{-21} &  3.3 \hspace*{1mm}10^{-20} & 16.5\\
\hline
\hline
\end{array}$
\label{min}
\end{table}
In Tables \ref{min} we have used the Schmidt nuclear model to allow easier comparison with previous work. But our conclusions are general, in particular the factors C/P is independent of the nuclear model used.

\subsection{Discussion of the improved nuclear model}
Recently new nuclear models beyond the Schmidt model for several atoms used in LIV experiments were published \cite{Brown2016, Flambaum2016}. We present here a different nucleus model described in Section \ref{sec:alt-nuc}. As discussed in Section \ref{sec:alt-nuc}, the results of our calculation differ significantly from the results for Cs presented in \cite{Flambaum2016}. When applying our method to the atoms also calculated in \cite{Brown2016} we find reasonable agreement. We also note that our results are qualitatively agreeing with the expectations from the Schmidt model, in so far as  $\gamma^n$ is a factor 45 smaller than $\gamma^p$ meaning that sensitivity to proton coefficients is much larger than to neutron ones, as expected. It is beyond the scope of this paper to analyse the different results in detail, here we simply remark that nuclear structure calculations are complex and thus the results have to be handled with care at this stage. We will provide more details of our calculations and their comparisons to other results in a future publication. 

Our results with the SCRMF (Table \ref{coeff-SCRMF}) set limits on the linear combinations $\ct^p_\mn + 0.021\ \ct^n_\mn$ of SME coefficients. For other atoms, e.g. those used in comagnetometers \cite{Smiciklas2011}, the corresponding linear combinations are different. This opens the possibility of combining different results in order to derive independent constraints on proton and neutron parameters in global analyses. This will be also addressed in more detail in future work.

\section{Conclusion and perspectives}\label{sect:conclusion}


We have presented new constraints on coefficients parametrizing Lorentz violations for nucleons in the minimal Standard-Model Extension (SME), by monitoring the frequency shifts of hyperfine transitions in a cold atom fountain on Earth. Within the fermion sector, our observable is by construction only sensitive to the $\cb_\mn$ coefficients (Sec. \ref{sec:lab-frame}). This test relies on the anisotropy of the kinetic energy in the wavefunctions of the nucleons, characterized by their non-zero momentum quadrupole moment whose value is highly nuclear model dependent. We first use the Schmidt model which has so far been mostly used (Sec. \ref{res_schmidt}), and extend the analysis to a more advanced SCRMF nuclear model (Sec. \ref{sec:alt-nuc} and \ref{sec:alt_res}). 

We have re-analysed data taken by the dual cold atom fountain FO2 at
SYRTE in $^{133}$Cs spin polarized mode on the $\ket{ 3 , m_F }
\longleftrightarrow \ket{ 4 , m_F }$ hyperfine transitions, as first
reported in \cite{Wolf2006}. Our analysis features the use of a new
SME model that includes terms of order $O\left( \beta^2 \right)$
(Sec. \ref{sec:SCF}). This allows access to the isotropic coefficient $\ct_{\text{{\tiny  \emph{TT}}}}$, not constrained by the previous analysis. Using a direct and simultaneous fitting of all parameters, we provide a complete analysis including individual limits on all $\ct_\mn$ coefficients, the associated correlation matrix, and confidence intervals (Sec. \ref{data_analysis}). We also present a description of the results in terms of independently constrained linear combinations, obtained through a singular value decomposition of the covariance matrix (App. \ref{svd}). 

 To allow for direct comparisons with previous works we have presented both results, based respectively on the Schmidt nuclear model (Table \ref{coeff}) and on our advanced nuclear model (Table \ref{coeff-SCRMF}). The present best limit on the $\ct_{\text{{\tiny  \emph{TT}}}}^p$ coefficient is improved by 5 orders of magnitude down to the $10^{-16}$~GeV scale (Sec. \ref{res_schmidt}). The advanced nuclear model allows to place limits on a linear combination of proton and neutron coefficients, in constrast to the Schmidt model used in \cite{Wolf2006} which accounts only for the proton sector sensitivity. This leads to improvements on the limits of the $\ct_{\text{{\tiny  \emph{Q}}}}^n$, $\ct_{\text{{\tiny  \emph{TJ}}}}^n$ and $\ct_{\text{{\tiny  \emph{TT}}}}^n$ neutron coefficients, by up to 12 orders of magnitude.

All our results are compatible with the absence of Lorentz violation. As mentioned in the introduction, if the relevant scale unit for suppressions is the electroweak-to-Planck energy ratio ($E_{ew}/E_P \sim 10^{-17}$),  it is particularly interesting to constrain the dimensionless LIV tensors $\ct_\mn/(mc^2)$ below this Planck suppressed scale. With this study we bring the limits at or below that level for the first time for all proton and neutron coefficients (with $mc^2\sim1$~GeV for protons and neutrons), except for $\cb_{\text{{\tiny  \emph{TT}}}}^n$ and $\cb_{\text{{\tiny  \emph{TT}}}}^p$ whose limits are respectively weaker by three and one orders of magnitude. Nonetheless, our results give an experimental benchmark indicating that the minimal suppression compatible with our data for the operators associated with $\cb_\mn$ is at least one Planck scale. These dimension 4 Lorentz violating terms could be expected to be of order one from Quantum Field Theory and the experimental observations thus require that an additional suppression mechanism comes into play, such as proposed in \cite{Liberati2013,Pospelov2012,Belenchia2016}.

Our $\ct_\mn$ coefficients estimations are still significantly correlated (see Fig. \ref{fig2}), mostly because our two data sets do not span a sufficient portion of the year to allow their full decorrelation from the annual sidebands (App. \ref{model}). Thus we expect that an additional data set would reduce the marginalized uncertainties and lead to an improvement by one extra order of magnitude of all limits, in particular bringing the limit on $\ct_{\text{{\tiny  \emph{TT}}}}^p$ down to $10^{-17}$~GeV. Based on synthetic data simulations, the best period for a third data set would be July or January. 

The $^{133}$Cs fountain data could also be analysed in an extended framework : the non-minimal SME framework, which takes into account higher order Lorentz violating operators. A non-minimal model, up to order 5 or 6, would contain additional sidereal harmonics, allowing for additional parameters to be determined and possibly for better decorrelation of the minimal ones (\cite{Kosteleck`y2015}).

Previous experiments realized with other set-ups and atoms could be reanalyzed with our new nuclear model. Analyzing SME spectroscopy experiments with more accurate models of the nucleus is a current effort of several groups (\cite{Flambaum2016,Brown2016}), and will be the subject of future work centered on the nuclear calculation method used here, and its application to other atoms and SME tests.

In conclusion, our work brings significant improvement in constraining possible low-energy signatures of new high energy physics, using a set of improved models and analysis that could also benefit to other experimental tests.

\acknowledgments
H.P.L.B. is grateful for the financial support of the Labex First-TF and the Centre national d'études spatiales (CNES). C.G., Q.G.B., and P.W. acknowledge support from the Sorbonne Universités grant Emergence for the CABESTAN collaboration. Q.G.B. was supported during the preparation of this work by the National Science Foundation under grant number PHY-1402890. C.G. would like to acknowledge useful discussions with A. Roura.

\appendix

\section{Second order model}\label{model}
\begin{table}
\caption{Composition of the complete $O\left( \beta^2 \right)$ model
  applied to $\ct_q$ in Eq. (\ref{eq-nuc}). The expression of the signal associated to each SCF coefficient is detailed in terms of frequency, phase and boost suppression. The only information not reported here are prefactors of order one from two angles, the lab colatitude and the inclination of the Earth's orbit. $\omega$ (resp. $\mathrm\Omega$) denotes the angular frequency of the Earth's rotation (resp. of the Earth's orbit). The column on the left is the main harmonic, and the column next to it is the sideband frequency. The spectral components included in the shortened model fitted to the data are in bold type.}
\flushleft
\scriptsize
$\begin{array}{c|c|c@{\,\vrule width 0.7pt\,}c|c|c|c|c|c|c|c|c}
    \hline\hline
   \multicolumn{3}{c@{\,\vrule width 0.7pt\,}}{\text{Frequency}} & \ct_{\text{{\tiny  \emph{Q}}}} & \ct_{-}& \ct_{\text{{\tiny  \emph{X}}}} & \ct_{\text{{\tiny  \emph{Y}}}} & \ct_{\text{{\tiny  \emph{Z}}}}&  \ct_{\text{{\tiny  \emph{TX}}}}& \ct_{\text{{\tiny  \emph{TY}}}}&  \ct_{\text{{\tiny  \emph{TZ}}}}&  \ct_{\text{{\tiny  \emph{TT}}}}\\
 \hline\hline

    \multirow{7}*{0} &   \multicolumn{2}{c@{\,\vrule width 0.7pt\,}}{\multirow{3}*{0}} &	\mathbf{1}&\beta_{\oplus}^{2}&\beta_\oplus^2&		&		&		&		&		&		\boldsymbol{	\beta_\oplus^2}\\
  &\multicolumn{2}{c@{\,\vrule width 0.7pt\,}}{  }&    \beta_\oplus^2&&		&		&		&		&		&&\boldsymbol{\beta_l^2}\\
&\multicolumn{2}{c@{\,\vrule width 0.7pt\,}}{  }&     	\beta_l^2	&	 &&		&		&		&		&		&\\
   \cline{2-12}
 &\multirow{2}*{$\mathrm{\Omega}$} &  \cos		 &		&		&	\beta_\oplus \beta_l	&	\beta_\oplus \beta_l	&		&		\boldsymbol{\beta_\oplus}&	\boldsymbol{	\beta_\oplus}&\boldsymbol{\beta_\oplus} &		\\
  &					&  \sin		 	&		&		&	\beta_\oplus \beta_l&	\beta_\oplus \beta_l	&		&		\boldsymbol{\beta_\oplus}&	\boldsymbol{	\beta_\oplus}&\boldsymbol{\beta_\oplus} &	\\
   \cline{2-12}
  &  \multirow{2}*{$ 2 \mathrm{\Omega}$} &  \cos		 &	\beta_\oplus^2	&	\beta_\oplus^2&	\beta_\oplus^2&	\beta_\oplus^2	&	\beta_\oplus^2	&		&		&		&\boldsymbol{	\beta_\oplus^2}\\
    &								&  \sin			 &	\beta_\oplus^2	&		\beta_\oplus^2&	\beta_\oplus^2	&		\beta_\oplus^2&	\beta_\oplus^2&		&		&		&	\boldsymbol{\beta_\oplus^2}\\
   \hline\hline
    \multirow{12}*{$\omega$} & 	 \multirow{2}*{$-2\mathrm{\Omega}$}	 &  \cos		 &	\beta_\oplus^2	&		\beta_\oplus^2&	\beta_\oplus^2	&		\beta_\oplus^2&	\beta_\oplus^2&		&		&		&\boldsymbol{\beta_\oplus^2}\\
    							    						&&  \sin		 &	\beta_\oplus^2&	\beta_\oplus^2&	\beta_\oplus^2&	\beta_\oplus^2	&		\beta_\oplus^2&		&		&		&	\boldsymbol{\beta_\oplus^2}\\ 
    							    						\cline{2-12}    							    						
    										  &  \multirow{2}*{$	-\mathrm{\Omega}$}		 &  \cos		 &\beta_\oplus \beta_l&\beta_\oplus \beta_l&\beta_\oplus \beta_l&\beta_\oplus \beta_l		&		\beta_\oplus \beta_l&	\boldsymbol{\beta_\oplus}&	\boldsymbol{	\beta_\oplus}&\boldsymbol{\beta_\oplus} &\boldsymbol{\beta_\oplus \beta_l}\\
    							    				&	 &  \sin		 &	\beta_\oplus \beta_l	&	\beta_\oplus \beta_l	&	\beta_\oplus \beta_l	&\beta_\oplus \beta_l&\beta_\oplus \beta_l&	\boldsymbol{\beta_\oplus}&	\boldsymbol{	\beta_\oplus}&\boldsymbol{\beta_\oplus} &	\boldsymbol{\beta_\oplus \beta_l}	\\
    							    				\cline{2-12}    
    										  &   \multirow{4}*{	0	} 	 &  \cos		 &		&		&		&	\mathbf{1}	&\beta_\oplus^2&		&\boldsymbol{	\beta_l}&		&		\\
    				&  & 	 &		&		&		&\beta_\oplus^2	&&		&&		&		\\
    							    				&		&  \sin		 &	\beta_\oplus^2&	\beta_\oplus^2&	\mathbf{1}&		&		&\boldsymbol{\beta_l}&		&		&	\boldsymbol{\beta_\oplus^2}\\
    							    				&		& 	 &	&&	\beta_\oplus^2	&		&		&&		&		&\\
    							    				\cline{2-12}    
     										  & 	 \multirow{2}*{$+\mathrm{\Omega}$	} 		 &  \cos		 &\beta_\oplus \beta_l&\beta_\oplus \beta_l&\beta_\oplus \beta_l&	\beta_\oplus \beta_l	&\beta_\oplus \beta_l	&\boldsymbol{\beta_\oplus}&	\boldsymbol{	\beta_\oplus}&\boldsymbol{\beta_\oplus} &\boldsymbol{\beta_\oplus \beta_l}\\
    							    				&	 &  \sin		 &	\beta_\oplus \beta_l	&		\beta_\oplus \beta_l&	\beta_\oplus \beta_l	&\beta_\oplus \beta_l&\beta_\oplus \beta_l&\boldsymbol{\beta_\oplus}&	\boldsymbol{	\beta_\oplus}&\boldsymbol{\beta_\oplus} &	\boldsymbol{\beta_\oplus \beta_l}	\\
    							    			\cline{2-12}  
      										  & 	 \multirow{2}*{$+2\mathrm{\Omega}$}		 &  \cos		 &	\beta_\oplus^2	&		\beta_\oplus^2&		\beta_\oplus^2&		\beta_\oplus^2&	\beta_\oplus^2&		&		&		&	\boldsymbol{\beta_\oplus^2}	\\
    							    						&&  \sin		 &	\beta_\oplus^2&	\beta_\oplus^2&	\beta_\oplus^2&	\beta_\oplus^2	&	\beta_\oplus^2	&		&		&		&	\boldsymbol{\beta_\oplus^2}\\ 
    																	   \hline\hline
    \multirow{14}*{$2\omega$} & 	 \multirow{2}*{$-2\mathrm{\Omega}$}	 &  \cos		 &\beta_\oplus^2&\beta_\oplus^2&\beta_\oplus^2&	\beta_\oplus^2	&	\beta_\oplus^2	&		&		&		&	\boldsymbol{\beta_\oplus^2 } \\
    							 &		&  \sin		 &	\beta_\oplus^2	&	\beta_\oplus^2	&\beta_\oplus^2		&\beta_\oplus^2&\beta_\oplus^2&		&		&		&\boldsymbol{\beta_\oplus^2}\\
    							\cline{2-12}   
    										  &  \multirow{2}*{$	-\mathrm{\Omega}$}		 &  \cos		 & & 		\beta_\oplus \beta_l	&		\beta_\oplus \beta_l	&	\beta_\oplus \beta_l&\beta_\oplus \beta_l		&\boldsymbol{	\beta_\oplus}&\boldsymbol{\beta_\oplus}  &		&		\\
    							    				 &		 &  \sin		 &&\beta_\oplus \beta_l&	\beta_\oplus \beta_l&			\beta_\oplus \beta_l&		\beta_\oplus \beta_l	&\boldsymbol{	\beta_\oplus}&\boldsymbol{\beta_\oplus} &		&			\\
    							    				\cline{2-12}    
    										  &   \multirow{6}*{	0	} &  \cos		 &	\beta_\oplus^2	&\mathbf{1}&	\beta_\oplus^2	&		&		&		&		&		&\boldsymbol{\beta_\oplus^2}	\\
    										  && 	 &	& \beta_\oplus^2	&&		&		&		&		&		&		\\
    										  &&  	 &		& \beta_l^2		&	&		&		&		&		&		&	\\
    							    				 &		&  \sin		 &		&		&		&		\beta_\oplus^2&	\mathbf{1}	&		&		&		&	\\
    							    				  &		&  	 &		&		&		&		&	\beta_\oplus^2		&		&		&		&	\\
    							    				 &		&  	 &		&		&		&	& \beta_l^2		&		&		&		&	\\
    							    				\cline{2-12}   
     										  & 	 \multirow{2}*{$+\mathrm{\Omega}$	}  &  \cos		 & & 		\beta_\oplus \beta_l	&		\beta_\oplus \beta_l	&	\beta_\oplus \beta_l&\beta_\oplus \beta_l		&\boldsymbol{	\beta_\oplus}&\boldsymbol{\beta_\oplus} &		&		\\
    							    				 &		 &  \sin		 &&\beta_\oplus \beta_l&	\beta_\oplus \beta_l&			\beta_\oplus \beta_l&			\beta_\oplus \beta_l&\boldsymbol{	\beta_\oplus}&\boldsymbol{\beta_\oplus} &		&			\\
    							    				 \cline{2-12}   
      										  & 	 \multirow{2}*{$+2\mathrm{\Omega}$}		 &  \cos		&\beta_\oplus^2&\beta_\oplus^2&\beta_\oplus^2&	\beta_\oplus^2	&	\beta_\oplus^2	&		&		&		&\boldsymbol{	\beta_\oplus^2}\\
    							 &		&  \sin		 &	\beta_\oplus^2	&		\beta_\oplus^2&	\beta_\oplus^2	&\beta_\oplus^2&\beta_\oplus^2&		&		&		&			\boldsymbol{\beta_\oplus^2}\\
    																	   \hline\hline
    \multirow{4}*{$3\omega$}    &  \multirow{2}*{$	-\mathrm{\Omega}$}		 &  \cos		 &		&	\beta_\oplus \beta_l&		&		&		\beta_\oplus \beta_l	&		&		&		&		\\
    							    				&		 &  \sin		 &		&		\beta_\oplus \beta_l	&		&		&\beta_\oplus \beta_l&		&		&		&		\\
    							    				\cline{2-12}     
     										  & 	 \multirow{2}*{$+\mathrm{\Omega}$	} &  \cos		 &		&\beta_\oplus \beta_l&		&		&			\beta_\oplus \beta_l&		&		&		&		\\
    							    				&		 &  \sin		 &		&	\beta_\oplus \beta_l		&		&		&\beta_\oplus \beta_l&		&		&		&		\\
							    				 \hline\hline
   \end{array}$
   \label{final}
\end{table}
The direction and amplitude of the lab velocity with respect to the SCF are respectively denoted $\textbf{n}$ and $\beta =\frac{v}{c}$.
The Lorentz boost matrix $\Lambda$ of the lab with respect to SCF is : 
\begin{equation} \label{equ:Lambda1}
\Lambda =\begin{pmatrix}
\gamma & \gamma \beta \textbf{n}^{T}\\
\gamma \beta \textbf{n} & \ \ \mathbb{I}_3  + \left( \gamma - 1 \right) \textbf{n}\cdot\textbf{n}^{T}
\end{pmatrix}
\end{equation}
Expanding $\gamma = 1/ (\sqrt{1 - \beta^2})$ to the second order in $\beta$ gives :
\begin{align}
\Lambda &=\begin{pmatrix}
1 + \frac{1}{2} \beta^{2} & \beta\textbf{n}^{T}\\
 \beta \textbf{n} & \mathbb{I}_3  + \frac{1}{2} \beta^{2}\textbf{n}\cdot\textbf{n}^{T}
\end{pmatrix} \\
&=  \underbrace{\begin{pmatrix}
1 &\beta \textbf{n}^{T}\\
 \beta \textbf{n} & \mathbb{I}_3 
\end{pmatrix}}_{\Lambda^{(1)}} + \underbrace{\begin{pmatrix}
\frac{1}{2} \beta^{2} & 0\\
0 &  \frac{1}{2} \beta^{2}\textbf{n}\cdot\textbf{n}^{T}
\end{pmatrix}}_{\Lambda^{(2)}}\label{A3}
\end{align}
where $\Lambda^{(1)}$ is the first order boost and $\Lambda^{(2)}$ is the second order boost.\par 
Following \cite{Bluhm2003}, the transformation matrix  ${\mathfrak T}$ from the SCF to the lab frame is given by the product :
\begin{equation} \label{equ:T}
 {\mathfrak T} = \left( \begin{tabular}{cccc}
$ 1 $& $0  $ & $0$ & $0$\\
$ 0$ &$ $\\
$0$ & &$R $ &\\
$0$ & & &
\end{tabular}\right)\cdot \Lambda 
\end{equation}     
where $R$ is the rotation matrix describing the orientation of the lab frame's axes directions in the SCF.

The total boost is the sum of the orbital boost of Earth and the boost of the lab relative to the Earth, with amplitudes respectively $\beta_{\oplus}$ and $\beta_l$. As eccentricity of Earth’s orbit gives rise to a maximal deviation of $2\%$ of its mean value, it leads to subleading order corrections in our model and can be neglected as shown in \cite{Bluhm2003}. Earth’s orbit is thus taken as circular.

As already mentioned the second order boost is only necessary for
terms involving $\ct_{\text{{\tiny  \emph{TT}}}}$, as all other
components of $\ct_{\mn}$ are dominated by zero or first order terms
in $\beta$. However, it turns out (see Appendix \ref{annex_quentin})
that for the $\ct_{\text{{\tiny  \emph{TT}}}}$ component all $O\left(
  \beta^2 \right)$ terms are obtained from solely the first part of
(\ref{A3}), the second part giving rise to only $O\left( \beta^4
\right)$ terms. Note also, that consequently any other second order
effects (e.g. geodetic precession  \cite{Soffel2003}) can be neglected.

The transformation matrix ${\mathfrak T}$ we obtain is then used in (\ref{equ:transfo}) to relate the lab frame parameters to the SCF ones. Applying it to $\ct_q$ of equation \ref{eq-nuc} leads to a time varying signal in case any of the SCF $\ct_{\mn}$ coefficients is different from zero. The full explicit model is not given here because of its length. We summarize its spectral structure in Table \ref{final} with associated boost suppression factors for each SME coefficient. Restricting to the relevant terms used in the final adjustment, the shortened model is composed of an offset plus 12 frequency components, which amounts to a total of 25 quadratures. 

\section{Lab frame $\cb_q$ in terms of $\cb_{TT}$ to second order in boost}
\label{annex_quentin}
The (instantaneous) Lorentz transformation of co-vector components from the SCF to the laboratory frame is written
\beq
u_\mu = {\mathfrak T}^\Xi_{\pt{\Xi}\mu} u_{\tiny \Xi},
\label{LTcv}
\eeq
where capital Greek letters denote components with respect to the SCF. 
For the $ \cb_\mn$ coefficients in the lab frame, the transformation is
\beq
\cb_{\mu\nu} = {\mathfrak T}^\Xi_{\pt{\Xi}\mu}{\mathfrak T}^\Pi_{\pt{\Pi}\nu} \cb_{\tiny \Xi \Pi}.
\label{LTcc}
\eeq

We now adopt the special case of isotropic SCF coefficients $\cb_{\tiny \Xi \Pi}$, in matrix form (recall that  $\cb_{\tiny \Xi \Pi}$ is traceless):
\bea
\left(
\begin{array}{cccc} 
\cb_{\emph{{\tiny  TT}}} & 0 & 0 & 0 \\
0 & \frac 13 \cb_{\emph{{\tiny  TT}}} & 0 & 0 \\
0 & 0 & \frac 13 \cb_{\emph{{\tiny  TT}}} & 0 \\
0 & 0 & 0 & \frac 13 \cb_{\emph{{\tiny  TT}}} \\
\end{array} 
\right)
\label{isotropic}
\eea
and we focus on the quadrupole set of coefficients in the lab frame $\cb_q=\cb_{xx}+\cb_{yy}-2\cb_{zz}$.
Using the Lorentz transformation and equation (\ref{isotropic}), 
the lab frame coefficients can be written as
\bea
\label{cq}
  \cb_q    &=& \big[  {\mathfrak T}^T_{\pt{\Xi}x}  {\mathfrak T}^T_{\pt{\Pi}x} +  {\mathfrak T}^T_{\pt{\Xi}y}  {\mathfrak T}^T_{\pt{\Pi}y}\\
     && - 2  {\mathfrak T}^T_{\pt{\Xi}z}  {\mathfrak T}^T_{\pt{\Pi}z} \big] \cb_{\emph{{\tiny  TT}}}+\big[  {\mathfrak T}^J_{\pt{\Xi}x}  {\mathfrak T}^J_{\pt{\Pi}x}\nonumber\\
     &&  +  {\mathfrak T}^J_{\pt{\Xi}y}  {\mathfrak T}^J_{\pt{\Pi}y} - 2  {\mathfrak T}^J_{\pt{\Xi}z}  {\mathfrak T}^J_{\pt{\Pi}z} \big] \fr 13 \cb_{\emph{{\tiny  TT}}}.\nonumber
\eea
Thus we would need to know the spatial part of the Lorentz transformation to second order in the boost velocity to get the desired expression.
However, there is a quicker method.

We use the defining property of Lorentz transformations and the flat spacetime metric:
\beq
\et_\mn =   {\mathfrak T}^\Xi_{\pt{\Xi}\mu}  {\mathfrak T}^\Pi_{\pt{\Pi}\nu} \et_{\tiny \Xi\Pi}.
\label{eta}
\eeq
From this equation we can re-express the terms involving $ {\mathfrak T}^J_{\pt{\Xi}j}$ using only the $ {\mathfrak T}^T_{\pt{\Xi}j}$ terms.  
For instance if we pick $\mu=\nu=x$ then we have from \rf{eta}
\bea
1 &=&  {\mathfrak T}^\Xi_{\pt{\Xi}x}  {\mathfrak T}^\Pi_{\pt{\Pi}x} \et_{\tiny \Xi\Pi}
\nonumber\\
&=& - {\mathfrak T}^T_{\pt{\Xi}x}  {\mathfrak T}^T_{\pt{\Pi}x} + {\mathfrak T}^J_{\pt{\Xi}x}  {\mathfrak T}^J_{\pt{\Pi}x},
\label{eta-xx}
\eea
thus implying 
\beq
 {\mathfrak T}^J_{\pt{\Xi}x}  {\mathfrak T}^J_{\pt{\Pi}x} = 1 + \left[  {\mathfrak T}^T_{\pt{\Xi}x}\right]^2 
\label{identity}
\eeq
with similar identities holding for the other terms in the last line of \rf{cq}.
Using identities like \rf{identity} in equation \rf{cq} we arrive at the expression
\bea
\cb_q  &=& \frac 43 \cb_{\emph{{\tiny  TT}}} \big[  {\mathfrak T}^T_{\pt{\Xi}x}  {\mathfrak T}^T_{\pt{\Pi}x} +  {\mathfrak T}^T_{\pt{\Xi}y}  {\mathfrak T}^T_{\pt{\Pi}y}
- 2  {\mathfrak T}^T_{\pt{\Xi}z}  {\mathfrak T}^T_{\pt{\Pi}z} \big],
\label{cq2}
\eea
from which it should be clear that only the first order boost in $\Lambda$ is needed to find the time-dependence of $\cb_q$ in terms of $\cb_{\emph{{\tiny  TT}}}$. A similar method was used in \cite{Kostelecky2016}.

\section{Independent bounds on linear combinations of SME coefficients using singular value decomposition}\label{svd}
\begin{figure}[H]
\begin{flushleft}
\hspace*{-5mm}\includegraphics[width =.56\textwidth]{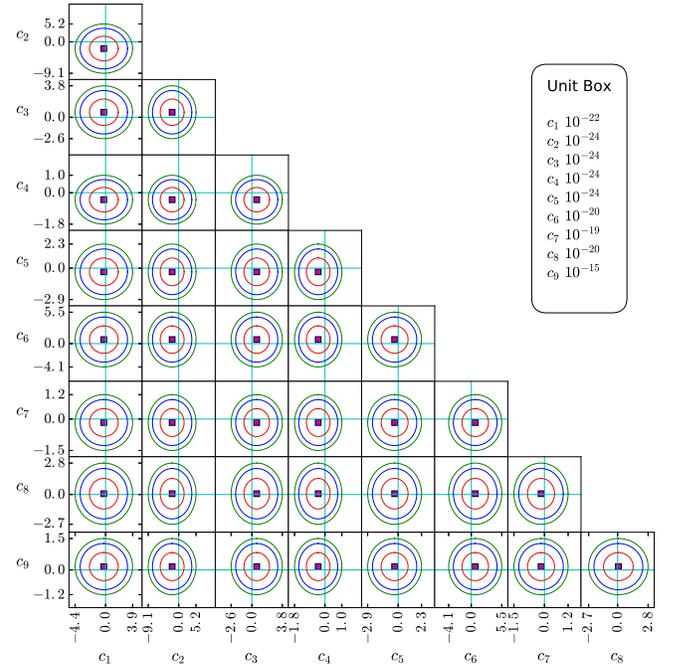}
\end{flushleft}
\caption{Confidence ellipses of the linear combinations (SVD). The red, blue and green confidence ellipses correspond respectively to the $ 68.3\%$, $90\%$ et $95.4\%$ confidence regions. The purple square is the position of the least-square solution. Axis labels give the $95.4\%$ confidence intervals in GeV, the respective orders of magnitude are given in the upper right box. Horizontal and vertical blue lines at 0 value allow to visualize the absence of significance in the results.}
\label{fig5}
\end{figure}

One can use a singular value decomposition of the covariance matrix \citep{numrec} to set constraints on linear combinations of SME coefficients. This decomposition is equivalent to a diagonalization of the covariance matrix $C$, which allows the determination of independently constrained linear combinations of parameters. These combinations are orthonormal to each other in the euclidean sense. We write the covariance matrix $C$ as
\begin{equation}
C =  V \cdot  S\cdot V^T
\end{equation}
where the matrix $V$ contains the eigenvectors of $C$ and the diagonal $S$ matrix contains its eigenvalues. These two matrices and the parameter vector $\textbf{a}$ are then used to define the linear independent combinations vector, denoted $\textbf{c}$, and their associated uncertainties $\sigma^{c}_{i}$:
\begin{align}
\textbf{c}& = V^T\cdot \textbf{a} \\
 \sigma^{c}_{i} & = \sqrt{ S_{ii}}. 
\end{align} 
The detailed composition of those combinations is presented in Table \ref{composvd} and their values and uncertainties are shown in Table \ref{coeffsvd}. We also provide the analytical confidence ellipses of these linear combinations in Figure \ref{fig5}. As expected, the ellipses are not tilted, meaning that the linear combinations have a diagonal covariance matrix. The results have been given for the Schmidt model. For the alternative nuclear model used here, results in Table \ref{coeffsvd} and Fig. \ref{fig5} would be rescaled as was done in Section \ref{sec:alt_res} for Table \ref{coeff-SCRMF}, and in their composition given in Table \ref{composvd}, proton coefficients would be replaced by $\ct_\mn^p+0.021\ct_\mn^n$.
\begin{table}
\caption{Limits on linear combinations of SME Lorentz violating parameters $\ct_\mn^p$ for the proton using the Schmidt model, in GeV. These combinations have been obtained using a singular value decomposition of the total covariance matrix.}
\centering
\begin{small}
$\begin{array}{c c c}
\hline \hline
\text{Linear combination}  &\text{ Value and uncertainty } &\text{ Unit (GeV)}\\
 \hline\vspace*{-3mm}\\ 
c_1 & (-0.2\pm 2.1 ) &10^{-22} \\
c_2 & (-2.0\pm 3.6) &10^{-24} \\
c_3 &  (0.6\pm 1.6) &10^{-24} \\
c_4 &  (-4.0\pm 7.0) &10^{-25} \\
c_5&   (-0.3\pm 1.3) &10^{-24} \\
c_6 &   (0.7\pm 2.4) &10^{-20} \\
c_7&  (-1.7\pm 6.8) &10^{-20} \\
c_8&  (0.1\pm 1.4) &10^{-20} \\
c_9 & (1.6\pm 6.9) &10^{-16} \\
\hline
\hline
\end{array}$
\end{small}
\label{coeffsvd}
\end{table}
\begin{table*}
\caption{Composition of the linear combinations of $\ct_\mn^p$ coefficients obtained using a SVD of the covariance matrix. Boldface numbers indicate the leading terms, for better visibility.}
\flushleft
$\begin{array}{c|ccccccccc}
\hline \hline
& \ct^p_{\text{{\tiny  \emph{Q}}}} & \ct^p_{-}& \ct^p_{\text{{\tiny  \emph{X}}}} & \ct^p_{\text{{\tiny  \emph{Y}}}} & \ct^p_{\text{{\tiny  \emph{Z}}}}&  \ct^p_{\text{{\tiny  \emph{TX}}}}& \ct^p_{\text{{\tiny  \emph{TY}}}}&  \ct^p_{\text{{\tiny  \emph{TZ}}}}&  \ct^p_{\text{{\tiny  \emph{TT}}}}\\
 \hline\vspace*{-4mm}\\ 

c_1 &\textbf{1.0}&	-5.3\hspace{2mm} 10^{- 5}&	3.5 \hspace{2mm}10^{- 5}&	-4.9 \hspace{2mm}10^{- 6}&	-7.7 \hspace{2mm}10^{- 6}&	-4.6 \hspace{2mm}10^{- 5}&	1.3 \hspace{2mm}10^{- 4}&	-1.7 \hspace{2mm}10^{- 5}&	2.8 \hspace{2mm}10^{- 9}\\
c_2 &5.7 \hspace{2mm}10^{- 5}&	\textbf{0.99	}&-0.01	&0.02 &	0.05 &	9.9 \hspace{2mm}10^{- 5}&	-8.1 \hspace{2mm}10^{- 5}&	4.2\hspace{2mm} 10^{- 5}&	-3.2 \hspace{2mm}10^{- 9}\\
c_3 &2.7 \hspace{2mm}10^{- 5}&	\textbf{-0.11}&\textbf{	-0.8}&	0.06&	\textbf{0.59}&	7.1\hspace{2mm} 10^{- 5}	&2.3 \hspace{2mm}10^{- 5}&	6.1\hspace{2mm}10^{- 6}	&-1.7\hspace{2mm}10^{- 9}\\
c_4 &-1.2 \hspace{2mm}10^{- 6}&	0.02	&-0.09&\textbf{	-0.99}&	-0.02&	-1.7\hspace{2mm} 10^{- 5}&	6.4 \hspace{2mm}10^{- 8}&	-1.4\hspace{2mm}10^{- 6}&	1.1\hspace{2mm} 10^{-10}\\
c_5 &1.3\hspace{2mm} 10^{- 5}&	-0.02&	\textbf{-0.57	}&0.06&	\textbf{-0.81}&	4.0 \hspace{2mm}10^{- 6}&	-1.8\hspace{2mm} 10^{- 5}&	3.7 \hspace{2mm}10^{- 6}&	-4.2 \hspace{2mm}10^{-10}\\
c_6&-6.3 \hspace{2mm}10^{- 5}&	1.1\hspace{2mm} 10^{- 4}&	-6.7 \hspace{2mm}10^{- 5}&	2.3\hspace{2mm}10^{- 5}&	3.9 \hspace{2mm}10^{- 5}	&\textbf{-0.96}&	\textbf{0.11}&	\textbf{-0.26	}&-7.1 \hspace{2mm}10^{- 5}\\
c_7 &1.1\hspace{2mm} 10^{- 4}&	-7.0 \hspace{2mm}10^{- 5}&	-9.8 \hspace{2mm}10^{- 6}&	2.2 \hspace{2mm}10^{- 6}&	2.9 \hspace{2mm}10^{- 5}	&\textbf{-0.22}	&\textbf{-0.86}&	\textbf{0.47	}&-2.1\hspace{2mm}10^{- 5}\\
c_8 &-6.1\hspace{2mm}10^{- 5}	&2.3 \hspace{2mm}10^{- 5}&	-2.3 \hspace{2mm}10^{- 6}&	2.5\hspace{2mm}10^{- 6}&-7.0\hspace{2mm} 10^{- 6}&	\textbf{-0.17}&\textbf{	0.51}&	\textbf{0.84	}&8.6 \hspace{2mm}10^{- 7}\\
c_9 &-4.8 \hspace{2mm}10^{- 9} &9.0\hspace{2mm} 10^{- 9}&	-6.8 \hspace{2mm}10^{- 9}	&2.0 \hspace{2mm}10^{- 9}&	4.2 \hspace{2mm}10^{- 9}&	-7.2\hspace{2mm} 10^{- 5}&	-1.1\hspace{2mm} 10^{- 5}&	-9.3\hspace{2mm} 10^{- 6}	&\textbf{1.0}\\

\hline
\hline
\end{array}$
\label{composvd}
\end{table*}

\section{Expectation value of the non-local $p^{2}$ operator}
Considering a generic one-body operator
\begin{equation}
\mathcal{\hat{O}} = \sum_{i,j} \braket{i|\mathcal{O}|j} c_i^\dagger c_j,
\end{equation}
one can show using the Wick theorem that the expectation value on an Hartree-Fock-Bogoliubov vacuum of this squared operator can be written:

\begin{multline}
\label{p2meanvalue}
\dfrac{\braket{\Phi_0|\mathcal{\hat{O}}^2|\Phi_0}}{\braket{\Phi_0|\Phi_0}}= \left[\text{Tr}(\mathcal{O}\rho)\right]^2 + \text{Tr}(\mathcal{O}^2\rho) \\ - \text{Tr}(\mathcal{O}\rho \mathcal{O} \rho) - \text{Tr}\left[\mathcal{O}\kappa(\mathcal{O}\kappa)^{*}\right],
\end{multline}
with the normal density defined as:
\begin{equation}
\rho_{ij} = \dfrac{\braket{\Phi_0|c_j^{\dagger} c_i|\Phi_0}}{\braket{\Phi_0|\Phi_0}},
\end{equation}
and the pairing tensor:
\begin{equation}
\kappa_{ij}= \dfrac{\braket{\Phi_0|c_j c_i|\Phi_0}}{\braket{\Phi_0|\Phi_0}}.
\end{equation}
In the case of \eqref{p2value} the matrix elements of the momentum operator can be expressed in the deformed harmonic-oscillator basis as:
\begin{equation}
\braket{i|\vec{p}|j} = - i \hbar \int d^3r_i d^3r_j \phi_i^\ast(\vec{r}) \vec{\nabla} \phi_j(\vec{r}).
\end{equation}
Thus one may directly use \eqref{p2meanvalue} to compute \eqref{p2value}.

\bibliographystyle{apsrev4-1} 
\bibliography{SMEclock1,SME1,divers1,nucleus1}

\end{document}